# The Turbulent 'Mixing' Layer as a Problem in the Non-equilibrium Statistical Mechanics of a Vortex Gas

**Saikishan Suryanarayanan**[1] **and Roddam Narasimha**[2]
*Jawaharlal Nehru Centre for Advanced Scientific Research,*
*Jakkur PO, Bangalore 560 064, India*

**N.D. Hari Dass**[3]
*Chennai Mathematical Institute, Chennai 603103, India*

# Abstract

The objective of this paper is to unravel any relations that may exist between turbulent shear flows and statistical mechanics, through a detailed numerical investigation in the simplest case where both can be well defined. The shear flow considered for the purpose is the 2D temporal mixing layer, which is a time-dependent flow that is statistically homogeneous in the streamwise direction ($x$) and evolves from a plane vortex sheet in the direction normal to it ($y$) in a periodic-in-$x$ domain with period $L$. The connections to statistical mechanics are explored by revisiting, via extensive computer simulations, an appropriate initial value problem for a finite but large collection of ($N$) point vortices of same strength ($\gamma$) and sign constituting a 'vortex gas'. Such connections may be expected to be meaningful as hydrodynamics, since the flow associated with the vortex gas is known to provide weak solutions of the Euler equation. Over ten different initial conditions classes are investigated using simulations involving up to $10^4$ vortices, with ensemble averages evaluated over up to $10^3$ realizations and integration over $10^4$ $L/\Delta U$ (where $\Delta U$ is the velocity differential across the layer, given by $N\gamma/L$).

The temporal evolution of the system is found to exhibit three distinct regimes. In Regime I the evolution is strongly influenced by the initial condition, sometimes lasting a significant fraction of $L/\Delta U$. Regime III is a long-time domain-dependent evolution towards a statistically stationary state via 'violent' and 'slow' relaxations (Chavanis 2011), over flow timescales of order $10^2$ and $10^4$ $L/\Delta U$ respectively. The final state involves a single structure that stochastically samples the domain, possibly constituting a 'relative equilibrium' (in the sense of Newton, pg.39).The distribution of the vortices within the structure is related to the Lundgren-Pointin (L-P) equilibrium distribution (with negatively high temperatures; L-P parameter λ close to −1), but has a non-isotropic truncated form because of the $x$-periodicity.

In-between is Regime II with a constant spreading rate, which is extensively studied in the turbulent shear flow literature as 'equilibrium', but is a part of the rapid non-equilibrium evolution which we label as 'explosive relaxation', and lasts less than $L/\Delta U$. The central finding is that this spreading rate is universal over the very wide range of cases considered here, with the value (in terms of momentum thickness) of $0.0166 \pm 0.0002$ times $\Delta U$. We find that Regime II is highly correlated with large values of $N$-independent two-vortex correlations and hence existing kinetic theories that neglect correlations or consider them as O($1/N$) would not be relevant for describing this regime.

The evolution of thickness in present simulations (in Regimes I and II) agree with the experimental observations of spatially evolving (3D Navier-Stokes) mixing layers, and the vorticity-stream-function relations in Regime III agree with those computed in 2D Navier-Stokes temporal mixing layers by Sommeria et al (1991). These findings suggest the dominance of what may be called the Kelvin / Biot-Savart mechanism in describing the large scale momentum and vorticity dispersal in the evolution of 2D turbulent free shear flows.

1. saikishan.suryanarayanan@gmail.com
2. roddam@jncasr.ac.in
3. dass@cmi.ac.in

# 1. INTRODUCTION

In a celebrated paper titled *Statistical Hydrodynamics*, Onsager (1949) presented a penetrating discussion of two-dimensional vortex dynamics in a 'gas' of positive and negative point vortices in an ideal fluid. (It is convenient to use the word 'gas' for describing the problem following Miller (1990), in spite of the fact that inter-vortex interactions described by the Biot-Savart relationship have very long range.) The motion of such a gas is governed by a Hamiltonian, and may be expected to lend itself to the formalism of statistical mechanics. (The demonstration that chaotic motion can occur in a collection of more than three vortices (Novikov & Sedov 1978, also see Aref 1983) establishes an underlying stochastic dynamics that justifies a statistical treatment.) Onsager showed that the motion of the vortices could be analyzed in terms of energy and entropy as in classical statistical mechanics, but the temperature derived therefrom would have to be permitted to take negative values, as the entropy has a maximum with respect to the energy. He also showed that such a gas possessed equilibrium solutions which consisted of large-scale vortex clusters or structures, positive and negative segregated from each other. Since then considerable work has been done in analyzing the mechanics of point vortices (see Paul Newton 2001 for example). In particular the nature of the equilibrium state in such a gas has been extensively discussed (see for example Lundgren & Pointin 1977, Eyink & Spohn 1993), especially in connection with the emergence of large-scale, long-lived vortices in the vortex gas. Several attempts (beginning with Marmanis 1997, most recently Chavanis 2010) have also been made to derive a BBGKY hierarchy of equations governing vortex distribution functions, based on the Liouville equation, beginning with single-particle analogues of the Boltzmann equation and followed by higher members in the hierarchy involving two-point correlations or more (Chavanis 2011). A favored target for application of these ideas has been Jupiter's famous red spot (Miller et al. 1992, Chavanis 2005), seen as one dramatic example of the kind of large-scale long-lived vortex predicted by Onsager.

In the fluid-dynamical literature, observations of large scale coherent vortical structures have been reported in many turbulent shear flows, including in particular the so-called 'mixing layer'. The spatially developing mixing layer (Figure 1a) is the flow which develops between two streams moving with different velocities $U_1$ and $U_2$, separated from each other for $x < 0$ by a thin splitter plate, and mixing with each other for $x > 0$. While extensive measurements have been made on this flow for more than fifty years (Liepmann & Laufer 1947, Brown & Roshko 1974, Winant & Browand 1974, Oster & Wygnanski 1982, Li et al 2009, Li et al 2010, Parezanovic et al 2012), the most striking development was the convincing demonstration by Brown & Roshko (1974) of the till-then unsuspected presence of highly organized large-scale vortices as an integral part of what was a canonical fully developed mixing layer flow. This work established that turbulent shear flows could contain ordered motion, and led to a search for and the study of such coherent structures in a wide variety of other shear flows (Liu 1989, Brown & Roshko 2012). The general point that all this work drove home was that the character of turbulent shear flows is fundamentally different from that of statistically homogenous isotropic turbulence, to the extent that ordered motion plays a significant (sometimes dominant) role in determining certain characteristics of sheared turbulence, such as for example entrainment of irrotational ambient fluid into rotational turbulent shear flow.

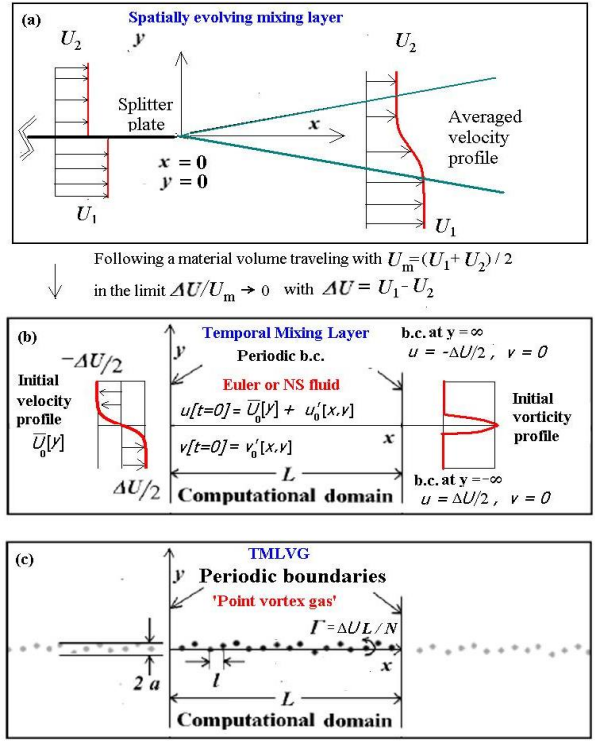

**Figure 1.a.** A schematic of a spatially evolving mixing layer. **b.** The temporal analogue (in an Euler or Navier-Stokes fluid) often studied in simulations. (Note that we use Reynolds decomposition, $\overline{U}$ indicates averaged velocity that depends only on $y$, and $u'$ and $v'$ are the $x$ and $y$ components of the fluctuating velocity which has zero mean. Subscript 0 indicates initial value). **c.** The vortex gas formulation of the temporal mixing layer showing the configuration of vortices at the initial instant. We track the vortices only in the $L$ – domain, $0 < x < L$ (which are denoted by dark dots). The governing equations account for the velocities induced by all the vortices in the $L$ - domain as well as all those present in $x < 0$, $x > L$ (shown in light colored dots) at separations of $+ kL$ and $- kL$ respectively ($k = 1,2, .. \infty$) for each vortex. $l = L/N$ is the initial inter-vortex separation in $x$.

The plane incompressible 'temporal' mixing layer (Figure 1b) is arguably the simplest conceivable turbulent shear flow, as its specification at high Reynolds numbers involves only one parameter, namely the velocity differential $\Delta U$ across the layer. This is a time-dependent flow that is statistically homogeneous in the streamwise direction $x$ and evolves temporally in the normal direction $y$, from an initial condition at $t = 0$ when the two streams moving at $+\Delta U/2$ and $-\Delta U/2$ are separated by a vortex sheet or a thin vortical layer at $y = 0$. The temporal mixing layer is related via a Galilean transformation to the spatially evolving case in the limit $\lambda \equiv 2(U_1 - U_2)/(U_1 + U_2) \to 0$. Further, it is favored for numerical simulations of the Navier-Stokes equations (e.g. Sommeria et al 1991, in 2D, Rogers & Moser 1994, in 3D), because of its simplicity and the unambiguous initial and boundary conditions that can be prescribed for the problem. It is usually studied in a domain $0 \leq x \leq L$ that is periodic in the flow direction $x$ with period $L$. This is a valid approximation to the infinite-domain mixing layer as long as the relevant length scales in the initial conditions and in the flow field are much smaller than the domain size.

The chief object of the present study is to make a comprehensive simulation of the vortex-gas analog of the temporal mixing layer, and to carry out a detailed statistical investigation of its evolution. As shown in Figure 1c, the problem can be formulated in terms of a row of equally spaced point vortices, located at $t = 0$ along the $x$-axis say, allowed to develop in the $x$-$y$ plane for $t > 0$. It is convenient and interesting to study the evolution for a class of initial conditions in which the vortices are given small displacements in *y* at *t* = 0. This system should not be viewed as a discrete model of a vortex sheet that rolls ups smoothly (for reasons that will be discussed in section 2 and Appendix C), but rather as a statistical (chaotic) evolution of a gas of point vortices.

The question of the possible relevance of point vortex dynamics to real world (3D Navier-Stokes) turbulent flows is a delicate one. There have been two major arguments against imputing such relevance. The first is the obvious one about dimensionality. There are however real-world flows which are quasi-two-dimensional in some sense: the most well-known of these is atmospheric motion at higher latitudes, where the large scales are governed by the dynamics of conserved potential vorticity oriented normal to the surface of the earth (Pedlosky, 1987). Indeed, the reverse energy cascade characteristic of 2D turbulence (Batchelor 1969, Kraichnan 1967) has given much insight into the dynamics of terrestrial and other planetary atmospheres. The second argument is about the complete absence of viscosity (and any molecular transport parameters that may be relevant for true mixing). No purely inviscid fluid can handle rigorously the phenomena of mixing and dissipation, both of which depend crucially on molecule-scale interaction, and consequently both the Richardson cascade and Kolmogorov-type similarity are beyond point vortex dynamics. Incidentally this implies that the word 'mixing' in characterizing the flow we study here is somewhat misleading. The vortex gas model does however describe what may be called the 'dispersal' of vorticity and hence also of momentum, both through the Biot-Savart relation. This is commonly, and from a fundamental point of view erroneously in turbulent flows (e.g. Narasimha 1990), thought of as a kind of 'diffusion', and often modeled through an 'eddy' viscosity, which in general is no more than an empirical convenience.

It has been argued that the long time evolution of vortex blobs in real fluids cannot be described by vortex gas motions as the effect of viscosity (say $v$), however small, does become manifest on time-scales of order $v^{-1}$. Interestingly, these arguments take on a different complexion in shear flows, especially in mixing layers. Plane turbulent mixing layers (2D in the mean) do have 3D structures and motions, but the large coherent structures that dominate the growth of the layer (in time or space) are quasi-two-dimensional (Brown-Roshko 1974, Wygnanski et al. 1979). One consequence of the streamwise/temporal growth in the size of the large-scale structures is that the local Reynolds number of the flow ($\delta \Delta U/v$, say, where $\delta$ is a measure of the layer thickness), actually increases linearly with downstream distance *x* in spatially evolving flow, and with time *t* in the temporally evolving flow that is the chief subject of the present study. Thus the effect of viscosity progressively diminishes (equivalently a locally scaled $v \to 0$) as $x \to \infty$ or $t \to \infty$, and the viscous timescale of O($v^{-1}$) consequently recedes to $\infty$ in the limit, *as long as the layer keeps growing*. In any case, some effects of viscosity can, if necessary, be taken into account by the addition of a random walk component in vortex motion (Chorin, 1973).

Another objection to the use of the vortex gas model is the singularity in the velocity field of the vortex gas at the location of the vortices. This can be overcome by desingularization of some kind (e.g. Krasny 1986), but it will be shown below that this does not affect our major conclusions and is unnecessary. Finally the temporal mixing layer does not have an equivalent of the feedback from downstream (which may become dominant if the velocity difference is comparable to the convection velocity) in spatially evolving mixing layers.

In spite of such objections, early vortex-gas simulations (Delcourt & Brown 1979, Aref & Siggia 1980) were remarkably successful in mimicking several dominant features of evolving mixing layers, such as the emergence and subsequent growth of the mixing layer through amalgamation events among the coherent structures. In retrospect they suffered from inadequate numerical accuracy in integration, small vortex populations and short integration times. Much more accurate and comprehensive simulations are however possible with today's computational resources and, regardless of any possible connection with real mixing layers, are of fundamental importance as a study of the simplest conceivable 'shear turbulence'. They further provide insights into understanding the interplay between chaos and order through a statistical-mechanics treatment of simple turbulent flows.

Our approach to the problem is akin to that of studying the statistical mechanics of a system of molecules via molecular dynamics. Usual molecular dynamics techniques become ineffective in the presence of long range forces; but in our context, the two-dimensionality of the problem somewhat compensates for this handicap. We therefore follow the complete evolutionary trajectory of the vortex-gas system all the way from its initial conditions (such as that shown in Figure 1c) to the final asymptotic state (if one exists) as $t \to \infty$. Compared to earlier work, the present simulations are much longer in time (by a factor of $10^4$), far more precise (Hamiltonian conserved to within $10^{-5}$), and involve large (500+ member) ensembles; these (as we shall demonstrate) turn out to be crucial for obtaining the results reported here.

The temporal development of the solution is analyzed from two view-points. The first is in terms of statistical mechanics, and describes the evolution of the vortex gas all the way from the initial condition to a final asymptotic state, through distribution functions, possible equilibrium states and temperatures. Such analyses point to the existence of certain universalities that appear to be novel in non-equilibrium statistical mechanics. The second viewpoint is in terms of concepts that have been found useful in the study of turbulent shear flows, such as self-similarity, growth rate of the mixing layer and effect of initial conditions on subsequent flow development. The two viewpoints together yield fresh insights into questions that have been widely discussed but remain controversial in the fluid-dynamical literature.

The organization of this paper is as follows. In section 2, we shall formulate the problem, present a critical review of earlier calculations and describe the present computational strategy. Then we shall discuss the results of our simulations, identifying and describing three distinct regimes in the temporal evolution in section 3; detailed results and analyses of the intermediate non-equilibrium universal regime (II) will be presented in sections 4 and 5. The domain-influenced regime (III) and the possible final asymptotic state of the system will be discussed in section 6. The relevance of the present study to Navier-Stokes mixing layers is described in section 7.

## 2. CURRENT APPROACH

### 2.1. Formulation

Many of the earlier vortex gas studies involve vortices in an infinite plane (e.g. Lundgren & Pointin 1977), in a doubly periodic box (e.g. Montgomery & Joyce 1974) or on a cylinder (e.g. Bühler 2002). The object of the present study is a temporal mixing layer in a point vortex gas, is formulated as follows.

Consider an array of *N* point vortices each of fixed strength $\gamma$, initially distributed along or very close to the *x*-axis (equispaced in *x*) in a domain of length *L* containing an inviscid fluid as shown in Figure 1c. Corresponding to any vortex *i* in the domain (say at $(x_i, y_i)$), there exist vortices at $\{(x_i + kL, y_i)\}$; $k = \{-\infty \ldots -2, -1, 0, 1, 2, \ldots \infty\}$, in the replicated domains extending to $\pm\infty$, so the boundary conditions are periodic. This formulation represents a canonical system of an infinite number of vortices in an infinite domain. Our objective is to study the evolution of this system in $(x,y,t)$ space.

The state of the system at any time *t* is completely described by the location of all the vortices $\{x_i[t], y_i[t]\}$, $i = 1$ to *N*, as this is sufficient to determine the velocity field over the whole domain. The velocity induced at a distance *r* by a point-vortex is given by the Biot-Savart relation

$$u_r = 0, \qquad u_\theta = \frac{\gamma}{2\pi r}, \qquad r > 0$$

where $u_r$ and $u_\theta$ are the radial and circumferential components of the velocity at the radial distance of $r = [(x - x_i)^2 + (y - y_i)^2]^{1/2}$.

The velocity with which any vortex moves is the vector sum of the velocities induced at its location by all the other vortices in the system; i.e. it is a flow-marker and traces a particle path. In the present set up, the velocity of a vortex located at $(x_i, y_i)$ in the *L*-domain is the sum of the velocities induced there by vortices at $\{(x_j + kL, y_j)\}$; $j = \{1 \text{ to } N\}, j \neq i; k = \{-\infty..-2,-1,0,1,2,..\infty\}$. This leads to convergent series that sum up to the following expressions for the *x* and *y* components of the velocity:

$$u_i = \frac{dx_i}{dt} = -\frac{\gamma}{2L} \sum_{j=1, j\neq i}^{N} \frac{\sinh(2\pi(y_i - y_j)/L)}{\cosh(2\pi(y_i - y_j)/L) - \cos(2\pi(x_i - x_j)/L)} \qquad (1)$$

$$v_i = \frac{dy_i}{dt} = \frac{\gamma}{2L} \sum_{j=1, j\neq i}^{N} \frac{\sin(2\pi(x_i - x_j)/L)}{\cosh(2\pi(y_i - y_j)/L) - \cos(2\pi(x_i - x_j)/L)} \qquad (2)$$

These equations appear to have been first written down by Friedmann & Poloubarinova (1928). The first reported calculations using (1,2) were performed by hand by Rosenhead (1931). Subsequent work using (1, 2) will be reviewed in Section 2.3.

We now set $\gamma = L\Delta U/N$, so that $\pm \Delta U/2$ are respectively the induced *x*-velocities at $y = \mp\infty$ as shown in Figure 1. It has to be noted that *x* is an angular variable as the system is *x*-periodic. In the numerical implementation, vortices that leave the domain during the evolution are relocated modulo *L* using the *x*-periodicity of the system. Thus if vortex *i* located at $x_i^m$ ($0 < x_i^m < L$) at time $t_m$ would have to be moved to $\hat{x}_i^{m+1} > L$ at time $t_{m+1}$, it is relocated to $x_i^{m+1} = \hat{x}_i^{m+1} - L$. Similarly, if $\hat{x}_i^{m+1} < 0$. Such operations do not alter the induced velocities given by (1, 2), and ensure that we always track *N* vortices all present within the domain $0 \leq x_i \leq L$.

The point vortex gas in an infinite plane possesses the Hamiltonian (Kirchhoff 1876)

$$\mathcal{H} = -\frac{\gamma^2}{4\pi} \sum_{i=1}^{N} \sum_{j=1; j\neq i}^{N} \ln(|\mathbf{r}_i - \mathbf{r}_j|/R_0) \qquad (3)$$

where $r_i \equiv \{x_i, y_i\}$ and $R_0$ is an arbitrary length scale, often taken as the radius of gyration of the vortex system, $[(1/N) \sum_{i=1}^{N}(x_i^2 + y_i^2)]^{1/2}$. For the system shown in Figure 1c, the Hamiltonian (often also called Kirchhoff's function) takes the form (Delcourt & Brown, 1979)

$$\mathcal{H} = -\frac{\gamma^2}{8\pi} \sum_{i=1}^{N} \sum_{j=1; j \neq i}^{N} \ln\left(\frac{1}{2}\left[\cosh(2\pi(y_i - y_j)/L) - \cos(2\pi(x_i - x_j)/L)\right]\right)$$

(4)

Equations (1) and (2) can be cast in the Hamiltonian form

$$\frac{d(x_i\sqrt{\gamma})}{dt} = \frac{\partial \mathcal{H}}{\partial(y_i\sqrt{\gamma})} \; ; \; \frac{d(y_i\sqrt{\gamma})}{dt} = -\frac{\partial \mathcal{H}}{\partial(x_i\sqrt{\gamma})}$$

(5)

a system of 2N ODEs that can be solved as an initial value problem. Note that in the present problem the notation L is used for the domain size and not the radius of gyration.

## 2.2. The major questions

Before posing the major questions some simple simulations over a relatively long duration are useful. These were performed with $N = 800$, initially equispaced in x and with initial values of the y-positions of the vortices from drawn randomly from a uniform probability distribution of amplitude $a$, $(P[y] = (1/2a) \text{ for } |y| < a \; ; 0 \text{ for } |y| > a)$.

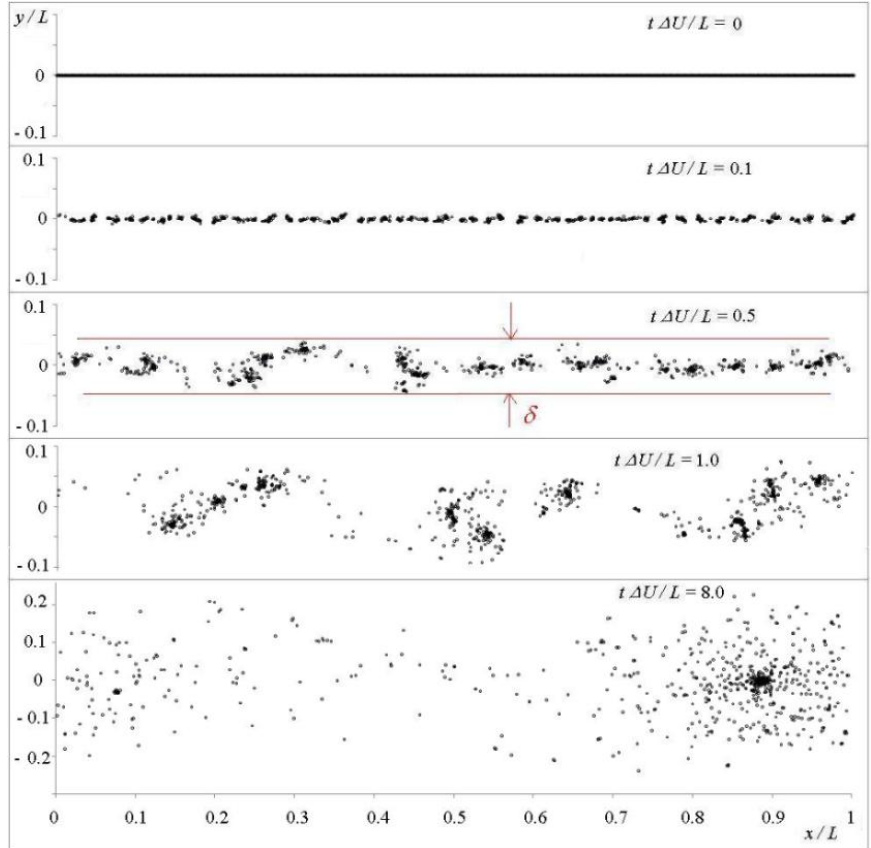

**Figure 2.** Typical evolution of vortex positions with time. ($N = 800$, $a/L = 10^{-6}$)

Figure 2 shows the evolution of vortex positions with time for $a/L = 10^{-6}$. The initial evolution is qualitatively consistent with earlier simulations of this kind (Delcourt & Brown 1979, Aref & Siggia 1982); in particular, as is clear from Figure 2 for example, the vortices cluster to form what has been called in fluid dynamical literature as 'coherent structures' that grow in size by successive amalgamations. The average size of the structures and the spacing between them increase with time, while the total number of structures in the domain decreases. We also find that beyond $t\Delta U/L \sim 4$, there is only one structure left in each periodic domain.

To quantify these observations, we introduce a rough measure of layer thickness $\delta(t)$, defined as the maximum $y$-distance at time $t$ between any two vortices in the system (see Figure 2). (This measure is analogous to the visual thickness of a laboratory mixing layer.) The evolution of $\delta$ with time is shown in Figure 3 for $a/L = 10^{-6}$ and $10^{-2}$. (Similar results are obtained if other measures of thickness are used instead of $\delta$.)

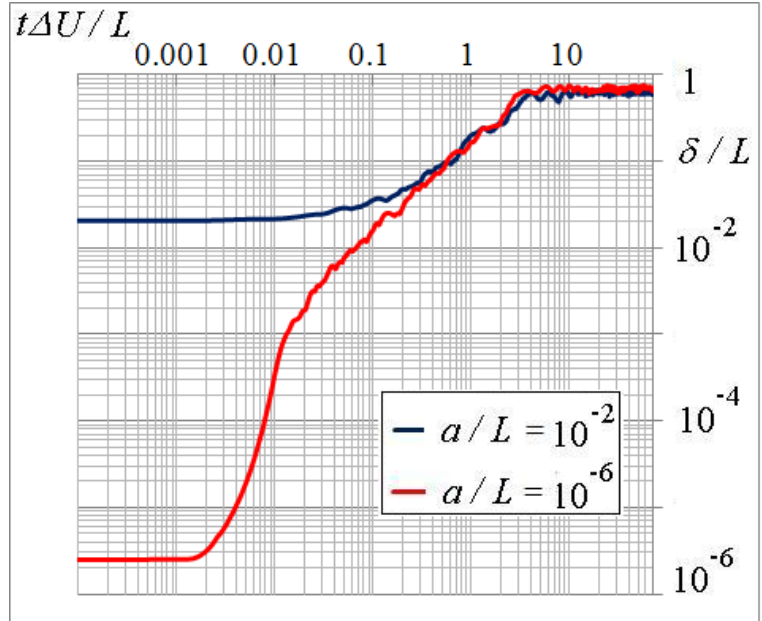

**Figure 3.** Evolution of thickness $\delta$ with time, in exploratory simulations of (1,2) (with $N=800$). Note the existence of different regimes in evolution.

It can be seen that (for $a/L = 10^{-6}$) $\delta$ grows approximately linearly between $0.04 < t\Delta U/L < 2$, saturating at about 0.8 for $t\Delta U/L \geq 4$. At the much higher initial amplitude $a/L = 10^{-2}$, the onset of linear growth takes place much later at around $t\Delta U/L \sim 0.2$, but the trajectory beyond that point seems to roughly follow the simulation with $a/L = 10^{-6}$, indicating a possibility of universal growth.

These two simulations immediately highlight the presence of at least three regimes in the evolution. For some time after initiation, the solution strongly depends on the initial condition (which we shall call Regime I), but the effects seem weaker at later times as $\delta$ grows linearly in time (Regime II). At longer times the layer thickness seems to fluctuate roughly around a constant value of about 0.8 (Regime III), which has never been explored in earlier simulations.

These preliminary simulations raise the following basic questions:

  a) What are the scaling laws in different regimes in the evolution of (1,2) ?
  b) Are any of the regimes 'universal', if so which ones and in what variables?
  c) Wherever there is universality, what is the statistical-mechanical explanation?
  d) What is the nature of the solution as $t \to \infty$ ?

We now briefly review earlier studies of vortex gas mixing layers with the above questions in mind.

**2.3. Review of Earlier Simulations**

Earlier vortex gas simulations (mostly carried with a fluid dynamical perspective) have not been explicitly identified the above three regimes and hence no attempt has been made to tackle the questions raised above. This is due to one or more of the following factors:

<u>a) Large statistical uncertainties due to small number of vortices, low accuracy and lack of ensemble averaging</u>

The most extensive vortex gas computations to date are due to Aref and Siggia (1982), who have $N = 4096$. They use a cloud-in-cell method which saves computer effort using integer algebra and look-up tables for the calculations, but the technique also introduces a numerical viscosity. With only a single realization they estimated the uncertainty level as 30% in the layer thickness. Delcourt & Brown (1979), also using a cloud-in-cell method, reported a 6% change in the Hamiltonian in the computations.

<u>b) Short integration times</u>

The maximum $t\Delta U/L$ reported in earlier work is 1.2 (Delcourt & Brown, 1979). This is grossly inadequate to reach an asymptotic state or even to uncover the different regimes observed in Figure 3.For certain classes of the initial condition, e.g. those involving long-wave sinusoidal displacements of the point vortices (e.g. Rosenhead 1931, Acton 1976), the time of integration is too short even to move out of Regime I.

<u>c)Desingularization</u>

It has been noted that point vortices are 'too chaotic' to provide a satisfactory discrete model for a vortex sheet (Hama & Burke 1960, Moore 1971, see also Leonard 1980). This difficulty may be overcome by adopting a desingularized version of (1,2) following Krasny (1986). Recent studies include Sohn (2005, 2010) and Abid & Verga (2011). Such desingularization is relevant to investigations on the smooth roll-up of a vortex-sheet but not for an inherently chaotic object like turbulent shear flow, desingularization suppresses chaos, delays transition to Regime II and does not affect the final conclusions; and the Hamiltonian (4) is no longer conserved (see Appendix C).

Some simulations (e.g. Sohn 2010) vary the number and strength of vortices in an adaptive fashion in order to better resolve the curvature of a continuum vortex sheet. The use of vortex sheet models for the mixing layer demand special techniques (e.g. Basu & Narasimha 1992, Paul & Narasimha 2012), but these are once again not attractive for statistical-mechanics approaches.

## 2.4. Present computational strategy

Here we start with $N$ vortices placed along $x$ with a given inter-vortex spacing $l$. At $t = 0$, they are displaced along $y$ by a specified amount. This displacement is typically randomly generated using a specified probability distribution for each case, but in a few special cases the $y$-displacement is a sinusoidal function of $x$. To obtain the time evolution (1,2) are solved numerically using a standard (explicit) fourth order Runge-Kutta algorithm to advance in time the locations of all the vortices. We perform all calculations in double precision. We adopt $\Delta t = 0.1\, l/\Delta U$ as the timestep and find that a reduction in $\Delta t$ by a factor of 4 does not materially affect conclusions on the evolution of averaged quantities, although individual vortex trajectories could be different due to the inherently chaotic nature of the system.

We do not adopt desingularization for the reasons highlighted in section 2.3. The conservation of the Hamiltonian prevents any two vortices from getting arbitrarily close to each other. We find that using the present algorithm and adopted time step, the distance a vortex moves during any time step rarely exceeds that to its nearest neighbor and is almost always at least an order of magnitude less and hence the unbounded velocity in the neighborhood of a point vortex does not present an issue in the numerical integration of (1,2).

The accuracy of the algorithm used here has been assessed in two ways. The first is based on computations on vortices in an infinite plane with similar initial conditions and parameters as in the $x$-periodic mixing layer. In this formulation, the $x$- and $y$-centroids, second moment and Hamiltonian are all conserved (see Newton 2001). In the computations the Hamiltonian is conserved to within of $9 \times 10^{-6}$ of its initial value at $t\Delta U/L = 0.78$ for $N = 3200$. The first moments of the vorticity distribution about the $x$ and $y$-axes are conserved to within $10^{-16}l$ and $3 \times 10^{-13}l$, and the second moment to within $1.3 \times 10^{-9}$, of their respective initial values. The second class of assessment is based on the solutions of (1,2) for the $x$-periodic system shown in Figure 1c. It was found that the Hamiltonian (4) was conserved to within $2.5 \times 10^{-5}$ of its initial value during integration over $t\Delta U/L = 0.75$, with $N = 3200$. (For other invariants of the present $x$-periodic system see Appendix A.)

Study of Regime III involves long-time integration so a shorter time step of 0.025 $l/\Delta U$ is adopted. As a result the Hamiltonian is conserved to within 0.5% for an integration time of 3.6 x $10^4 L/\Delta U$ (0.58x$10^9$ time steps). These numbers demonstrate that the current computations are substantially more accurate than any previous work.

Apart from $\delta$ (defined in Section 2.2) there are different metrics one can adopt to specify the 'thickness' of the layer such as moments of vortex $y$-positions. But in order to enable comparison with Euler & Navier-Stokes mixing layers, we adopt the so-called momentum thickness, as it is commonly used in the fluid dynamic literature and in several earlier vortex gas mixing layer studies (e.g. Aref & Siggia 1982). The momentum thickness ($\theta$) is defined in the usual way as,

$$\theta[t] = \frac{1}{4}\int_{-\infty}^{\infty} dy \left(1 - \left(\frac{\overline{U}[y,t]}{\Delta U/2}\right)^2\right), \tag{6}$$

where, the $x$-averaged $x$-velocity $\overline{U}[y,t] = (1/L)\int_0^L dx\ u[x,y,t]$ is computed by $x$-averaging the induced $x$-velocity $u$ on a grid of 0.4 $N$ points in $x$ and 200 points in $y$ once every 100 time steps. (Note: There are rare occasions when a vortex can come arbitrarily close to a grid point and induce very high velocity. This can reflect in the $x$-averaged velocities and hence the momentum thickness. In principle this effect can be addressed by use of a very fine grid and by averaging over a thin strip which would lead to cancelling of the large induced velocities of opposite signs. But we find that neglecting the contributions made by those rare instances when $x$-averaged velocities with absolute value greater than $\Delta U/2$ while computing $\theta$ is an equivalent alternative easier for numerical implementation. We note that this strategy does not change the computed value of $\theta$ by more than 1% for 99.9% of cases when $t\Delta U/l > 10$ suggesting $\theta$ to be a robust measure).

We shall show in section 4 that our conclusions are not affected by the choice of measure of thickness.

**2.5. Ensemble averaging**

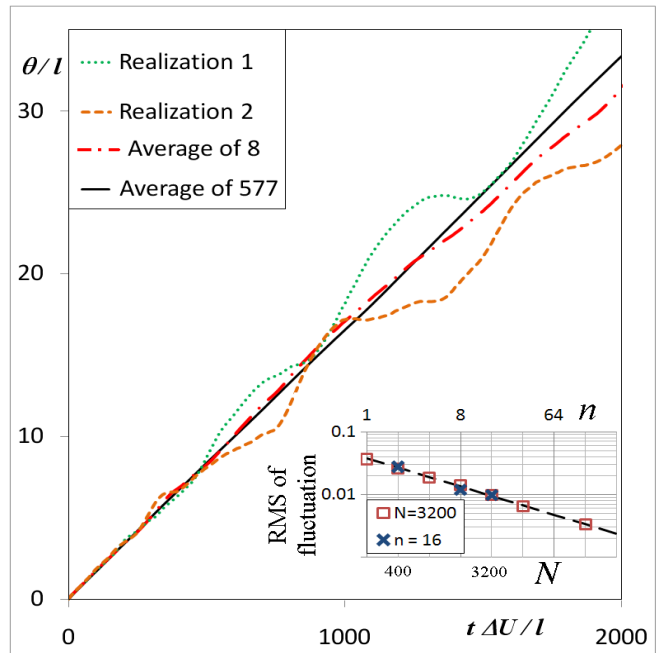

**Figure 4.** .Effect of ensemble averaging. Note that individual realizations have large fluctuations (even for $N = 3200$) and average over a large number of realizations is essential. The RMS departure from the respective means (at $t\Delta U/l = 160$) ~ $n^{-1/2}$ (shown in dashed line) for a given $N$ and ~ $N^{-1/2}$ for a given $n$.

In statistical mechanics ensemble averaging is commonly adopted to reduce fluctuations. For measurements of turbulent flow in fluid dynamics long-time averaging is often adopted as an alternative to ensemble averaging in statistically stationary flows. As the present system is non-stationary in time but statistically homogenous in $x$, $x$-averaging is in principle equivalent to ensemble averaging. However we find that an average over an ensemble of realizations (with initial conditions varied within a clearly specified class) is adopted due to the following reasons.

We first note that the statistical error (at a given $t\Delta U/l$) may be expected to vary with $N$ as $N^{-1/2}$ (observed also in our present simulations, as shown in Figure 4), while the computational effort grows as $N^2$. But if we simulate $n$ realizations with $m$ vortices each, the computational effort grows as $nm^2$ while the statistical error goes as $(nm)^{-1/2}$ (Figure 4). The errors will be the same if $N = nm$, which reduces the computational effort by $nm^2/N^2 = 1/n$. It is this result that makes the ensemble approach so attractive. But a sufficiently large *L/l* may be required to have a sufficiently long extent and preserve the inherent distinction between the different regimes observed in Figure 3. But once *N* is sufficiently large the ensemble averaging approach is computationally far more economical. It also has the practical advantage of using parallel computers more effectively, as different 'realizations' can be independently simulated on different processors without any need for data communication. We also find that the ensemble average of *θ* computed from the *x*-averaged velocity profile for each realization is not very different (for large *N* and *n*) from the value computed from the ensemble average of the (*x*-averaged) velocity. Throughout this study, we shall use the former for the sake of numerical convenience. We also note that, for a given initial condition class, the standard deviation of the Hamiltonian across realizations is never more than 1% of its mean value for present simulations with more than 400 vortices, and is often much less: e.g. less than 0.01% for the set of simulations presented in Section 4. Hence the present ensemble can be considered a microcanonical ensemble.

We shall discuss the significant implications of inadequate averaging in detail in Section 7.

### 3. RESULTS : THE THREE REGIMES IN EVOLUTION

With the objective of determining the precise scaling laws in each of the three regimes already noted in the simulations, we carry out several additional simulations with different initial displacements drawn from uniform random distributions with amplitudes ranging from $10^{-4} l$ to 10 *l*, with different domain sizes ranging from 200 *l* to 1600 *l*, and with averages over up to 12 realizations. A summary of the results is presented in Figure 5 as a composite diagram. In order to shed light on the different scaling laws in the different regimes, it is useful to adopt a measure of the layer thickness as *δ* or *θ* and of time as *l/ΔU* or *L/ΔU*. It is therefore important to pay attention to the precise variables used as abscissa and ordinate in Figure 5.

<u>Initial condition dominated Regime I.</u>

As Figure 5 displays, during the initial Regime I with time scaled with *l*, the evolution is widely different for different initial conditions. *δ* is adopted as the measure of layer thickness in this regime as *θ* is not accurate for *tΔU/l* < 10 as elaborated in section 2.4. It is seen that the duration of this regime ($t_{RI}$), $10^{-2}$ to 10 times $l/\Delta U$ for the cases considered here, strongly depends on the initial conditions as shown in an inset in Figure 5. For certain initial condition classes, including those where the *y*-displacement of vortices is a long-wave sinusoidal function of *x*, Regime I may be much longer (O($10^3$) $l/\Delta U$ for case P1 shown in section 4). In such cases the transition to Regime II may even be non-monotonic.

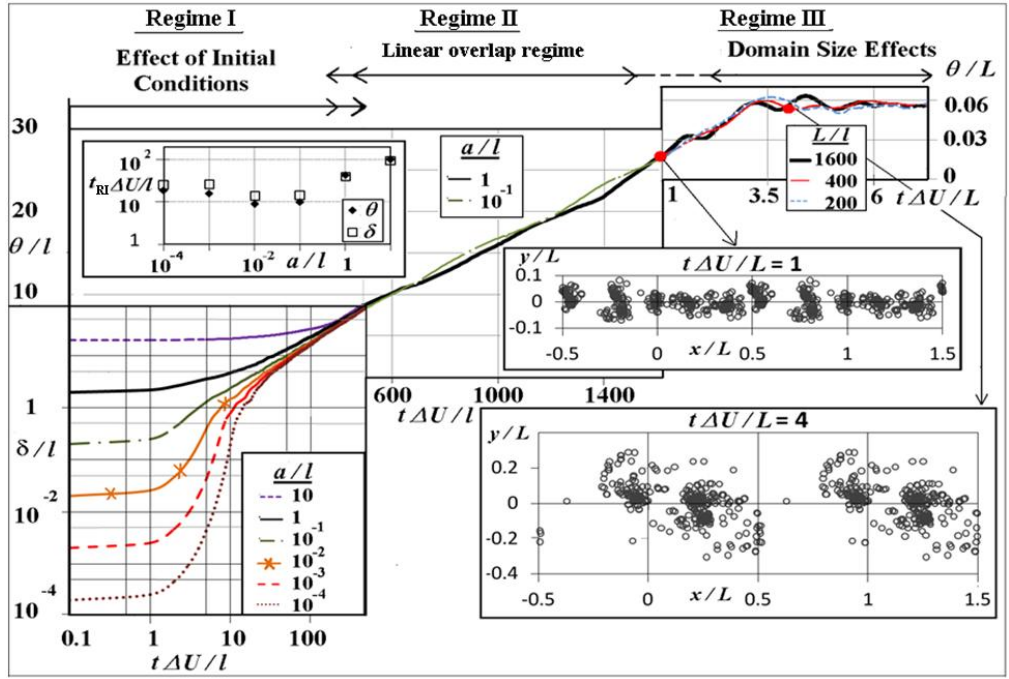

**Figure 5.** Composite diagram showing effect of initial conditions and domain size on the evolution of the mixing layer. Note use of $\delta$ and $\theta$ in different parts of the diagram, and change in the abscissa from $t\,\Delta U/l$ with a logarithmic scale up to 500, linear scale between 500 and 1500, and a switch to $t\,\Delta U/L$ thereafter. Appropriate changes have been made on both abscissa and ordinate to ensure that the evolution curve should go smoothly from one regime to the next. Inset on top left gives variation of initial transient with the amplitude of the initial vortex displacement. Two insets on the right give pictures of the configuration of the vortices at $t\Delta U/L = 1$ (upper) and at $t\Delta U/L = 4$ (lower).

Domain-limited Regime III

Jumping now to Regime III, we find from Figure 5 that, at times comparable to or larger than the domain size time-scale (i.e. $t\Delta U/L \geq 1$), the effects of finite domain size become noticeable and the growth of the layer departs from the linear variation with time seen in Regime II. As shown in the inset in Figure 5, the dynamics in the initial stages of Regime III are governed the interaction among a small number of coherent structures. Figure 5 shows that the scaling length clearly changes from $l$ to $L$ in this regime, as confirmed by the approximate collapse of $\theta/L$ vs. $t\Delta U/L$ obtained from simulations with $L/l$ ranging from 200 to 1600. Beyond $t\Delta U/L \sim 4$ the magnitude of changes in the thickness of the layer (in a statistical sense) is greatly reduced. This is because there is only one structure left in the domain (see lower inset in Figure 5), and hence there is no further opportunity for the layer to grow by amalgamation among structures. The evolution of the single structure to its final stage and its connections to vortex gas equilibrium are discussed in detail in Section 6.

The intermediate linear Regime II

It can be observed from Figure 5 that between Regimes I and III is an intermediate Regime II in which the layer exhibits linear growth.

From a mechanics view point, the transition between the short-time initial and long-time asymptotic states is governed by an intermediate asymptotics that can be derived by methods similar to those used by Millikan (1938) in channel flow and Kolmogorov (1941) in turbulence spectra (see Narasimha 1996). The argument can be applied to the present system as follows

Let $\hat{\delta}[t]$ be a measure of the mixing layer thickness uniquely determined from $\{x_i(t), y_i(t)\}$ at each instant of time $t$. The value of $\hat{\delta}$ could be a statistic directly involving the position of the vortices (such as the RMS value of $y$ - displacements of all vortices), or a thickness based on the computed $x$-averaged velocity field, such as the momentum or vorticity thickness. Dimensional analysis shows that the growth rate of the mixing layer can be written as

$$\frac{d\hat{\delta}}{d(t\Delta U)} = F\left[\frac{t\Delta U}{l}, \frac{L}{l}, \frac{\{x_i(0), y_i(0)\}}{l}\right] \qquad (7)$$

where $F$ is some function to be determined. (We choose $\Delta U$ as a basic variable instead of $\gamma$, as it is a large scale quantity that is more relevant for fluid dynamics, and this choice will not make any difference to the following analysis.) Here $l$ is the initial inter-vortex spacing, but it is important to note that the following analysis holds if $l$ were to be replaced by another characteristic length scale of the initial condition (such as amplitude or wavelength of periodic forcing). The limit $L/l \to \infty$ would imply that the domain size is much longer than any length scale characterizing the initial conditions. Such a limit would always be appropriate for any 'canonical' temporal mixing layer in an infinite domain.

If we hypothesize that the solution (7) evolves to a state independent of the precise initial configuration for sufficiently large $t\Delta U/l$, the third argument of $F$ in (7) will drop out in the limit, so

$$\lim_{t\Delta U/l \to \infty} \frac{d\hat{\delta}}{d(t\Delta U)} = \lim_{t\Delta U/l \to \infty} F_1\left(\frac{t\Delta U}{l}, \frac{L}{l}\right) = F_2\left(\frac{t\Delta U}{l}, \frac{L}{l}\right) \qquad (8)$$

where $F_2$ is the functional form taken by $F_1$ as $t\Delta U/l \to \infty$.

Assuming that this converges in the limit $L/l = N \to \infty$,

$$\lim_{N\to\infty, t\Delta U/l \to \infty} \frac{d\hat{\delta}}{d(t\Delta U)} = \lim_{N\to\infty} F_1\left(\frac{t\Delta U}{l}\right) = F_2\left(\frac{t\Delta U}{l}, N\right) \qquad (9)$$

In what may be called the long-time or 'outer' limit (see Van Dyke 1964), $t\Delta U/L = O(1)$, so the solution maybe expected to be dominated by the finite domain size and hence depend on $t\Delta U/L \equiv (t\Delta U/l) / (L/l)$, which we observe as we approach Regime III in Figure 5. This is known to happen in other areas of physics. Hence, in this regime, we may write (9) as

$$\frac{d\hat{\delta}}{d(t\Delta U)} \to F_3\left(\frac{t\Delta U}{L}\right), \quad t\Delta U/l \to \infty, L/l \to \infty, t\Delta U/L \text{ fixed}$$

(10)

where $F_3$ is the functional form assumed by (9) in the limits stated above. The above argument would strictly hold only in the early part of Regime III ($t\Delta U/L < 4$), as we shall show in Section 6.

If we postulate an overlap between (9) and (10) in the simultaneous limits $t\Delta U/l \to \infty$ and $t\Delta U/L \to 0$ (in the spirit of matched asymptotic expansions, Van Dyke 1946), the only possibility is an overlap Regime II in which

$$\frac{d\hat{\delta}}{d(t\Delta U)} = C_1 \qquad (11)$$

where $C_1$ is independent of time; i.e. the layer thickness grows linearly with time. This is the analog of the log law in channel flow and the $k^{-5/3}$ law in the spectrum. The question whether $C_1$ is universal (for different initial condition classes) will be addressed in Section 4.

## 4. UNIVERSALITY OF REGIME II

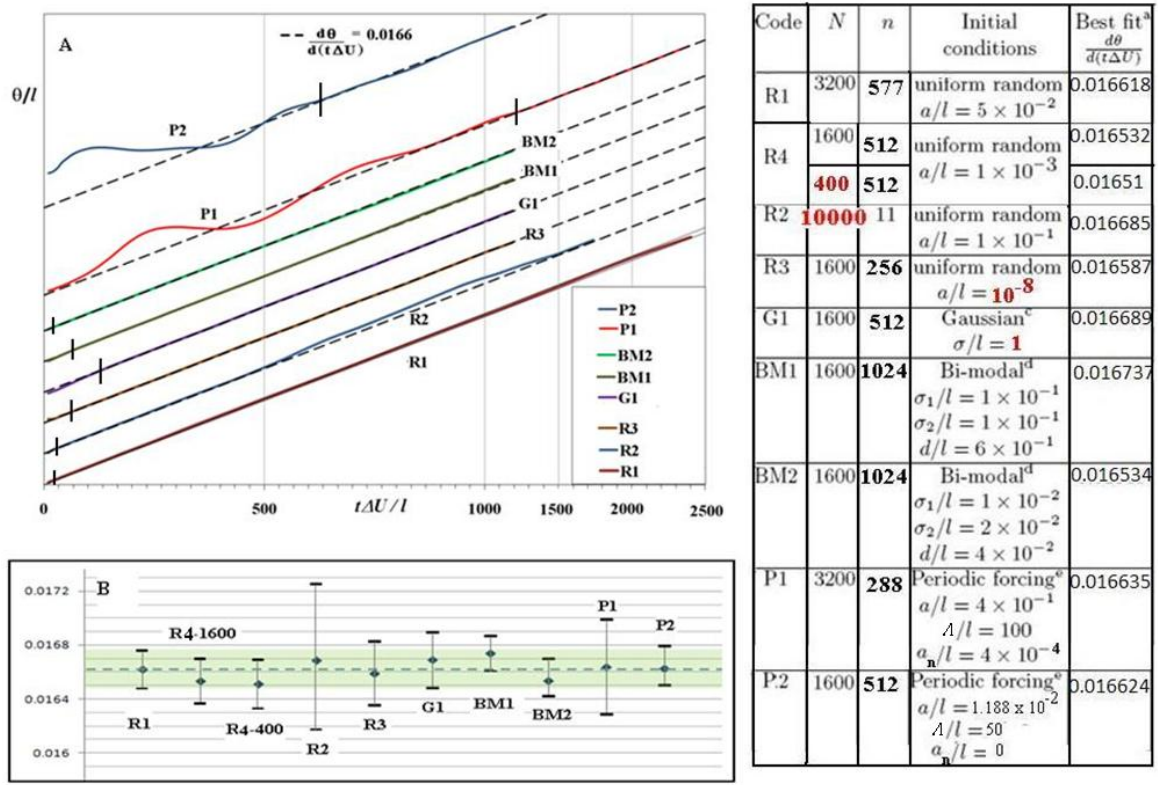

**Figure 6.** A. Universality of Regime II. Note the wide range of initial conditions including those with very long transients. See the change in scale beyond $t\Delta U/l$ of 1000. B. Estimate of uncertainties in Regime II growth rate. The error bars show the 95% confidence limits (computed using Student's t-distribution). The dotted line is drawn through the reference (R1) growth rate.

In order to test whether the Regime II growth rate is universal, a total of 10 cases, with widely different initial condition classes for $y_{i0}$ domain lengths, number of vortices and ensemble sizes have been performed. The results are presented in Figure 6.

The initial conditions considered include uniform random distributions (cases R1, R2, R3, R4-1600, R4-400) with amplitude ratio $a/l$ varying from $10^{-8}$ to $10^{-1}$; Gaussian distributions(G1);bi-modal distributions in the form of sums of symmetric or asymmetric displaced Gaussians (respectively BM1, BM2 ; the respective PDFs are shown in Appendix B); and distributions varying sinusoidally in *x* (P1 and P2). In the case of random initial conditions each realization is initialized with a different set of random numbers from the same class. The different 'realizations' required for ensemble averaging for sinusoidal initial conditions ($y_{oi} = a \sin(2\pi x_i/\Lambda + \phi_o)$, where *a* and *Λ* are the amplitude and wavelength of the perturbation) can be generated with different initial phases ($\phi_o$) of the wave at $t = 0$ with respect to that at the beginning of the domain ($x = 0$); small differences in numerics lead to different solutions in terms of evolution of individual vortex positions over time due to the chaotic nature of the system, but not in the statistics. This strategy is used in case P2. An alternative is to add a small random noise component to the wave at the initial instant, and draw it from some specified distribution. This is done for case P1 whose discussion is deferred to Section 7.

The respective growth histories in Regime II are shown in Figure 6. A best fit to the growth is obtained by minimizing

$$\sum_{t=t_{IIb}}^{t=t_{IIe}} (1 - ((At\Delta U + B)/\theta[t]))^2$$

with respect to *A* and *B*, where $t_{IIb}$ and $t_{IIe}$ are the estimated beginning and end of Regime II. We choose $t_{IIe}$ to be 0.8 $t\Delta U/L$ or the end of the simulation, whichever is earlier. The locations of $t_{IIb}$ are indicated in Figure5 by short vertical bars. We take as reference the best fit value for R1 ($N = 3200$; $n = 577$), in which Regime II extends over more than two decades in $t\Delta U/l$ (20 – 2400), and $d\theta/dt = 0.0166\ \Delta U$ + const.

Figure 6B shows the ensemble-averaged best-fit growth rates and the 95% confidence limits for the ten cases considered. Based on these results the evolution of momentum thickness in Regime II is given by

$$\theta_{\text{Regime II}} = 0.0166\ [\pm 0.00015]\ t\Delta U + C_3 \qquad (12)$$

with a universal slope and a non-universal intercept ranging from $-3.1\ l$ (P2) to $0.7\ l$ (G1) in the present simulations (the corresponding 'virtual origin' (intercept on the time axis) of the linear growth in Regime II, $t_0$, would be $186\ l/\Delta U$ and $-42\ l/\Delta U$). The departures in Regime II growth rate across the wide range of initial conditions are within a band of $\pm 1\%$ from the reference, as compared to the 30% uncertainty quoted by the authors in the vortex gas simulation of Aref & Siggia (1980)

Figure 7 shows that profiles of mean velocity and Reynolds shear stress (evaluated by the integration of vorticity flux (6)) for the case P1 and (at two different times) for the case G1. In the normalization used in similarity theory, with velocities scaled by $\Delta U$ and normal displacement with $\theta$, it is seen that the three profiles agree for both mean velocity and Reynolds stress, indicating both self-similarity and universality, and hence of (fluid-dynamical) equilibrium in the sense of Narasimha & Prabhu (1971).This implies that universality extends to any measure of thickness based on the mean velocity profile.

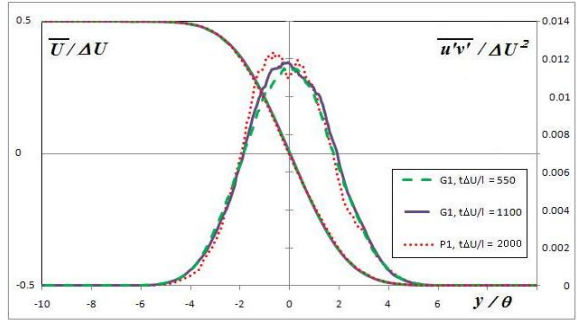

**Figure.7.** Self similarity and universality of *x*-averaged (fluid) velocity and 'Reynolds shear stress' profiles. The latter has been evaluated using integral of vorticity flux (computed from 64 member ensembles).

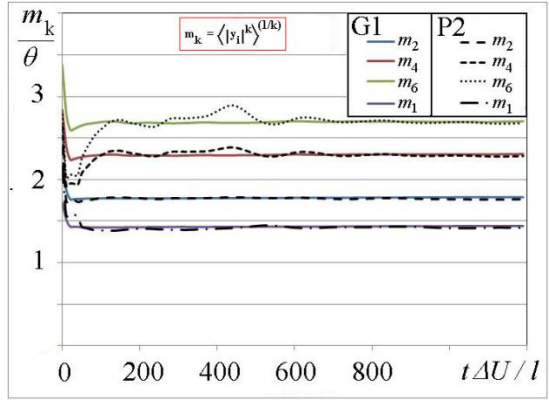

**Figure.8.** Evolution of various measures of thickness based on vortex positions for G1 and P2. All of them settle to a constant factor of $\theta$ in Regime II for the two very different initial conditions.

For two illustrative cases (G1 and P2) figure 8 shows that the moments of the vortex positions, $m_k = <|y_i|^k>^{(1/k)}$, become universal multiples of the momentum thickness at sufficiently long times, in general longer for the sinusoidal initial condition (P2) compared to the Gaussian initial condition (G1), establishing similarity and universality irrespective of the measure used to describe layer thickness.

## 5. THE NON-EQUILIBRIUM STATISTICAL MECHANICS OF REGIME II

Is there a statistical-mechanical explanation for the observed universality in Regime II ? To answer this we first explore the possibility of describing Regime II using approaches based on existing 'vortex gas kinetic theories' inspired by the Boltzmann equation (e.g. Marmanis, 1998, Chavanis 2001,2005,2010,2011, Sano, 2008). This may be done by computing the single- and two-vortex distribution functions in the present simulations. We consider cases R4-1600 and R4-400 as they involve a short Regime I and large ensemble sizes, and also provide an opportunity to assess the effect of the number of vortices on the simulations. We also analyze simulations with $a/l = 2$, $N = 400$ to study the effect of initial conditions (if any). The domain is divided into 40 by 40 boxes of equal size, the width of each box ($\Delta x$) being fixed, while the height ($\Delta y$) increases linearly with time to cover the entire layer at each instant with optimum resolution. The number of vortices present in each box at a given time is averaged over an ensemble of 512 realizations. This gives the single vortex distribution function $f_1$ (x, y, t), which is normalized such that $\sum f_1 \Delta x \Delta y = L^2$. (We use this normalization for convenience as renders $f_1$ dimensionless, in contrast with the conventional definition. Note that this single particle distribution function is related to the averaged vorticity as $f_1 = \langle \omega \rangle / (\Delta U/L)$) Since the system is homogeneous in x, large-ensemble averages should be independent of x. But even for the ensemble sizes used in the present simulations, there are fluctuations in $f_1$ of upto 10% along x. Therefore, in order to improve the statistics, $f_1$ (x, y, t) may be averaged over x to obtain $\overline{f_1}[y]$, with $\sum \overline{f_1} \Delta y = L$.

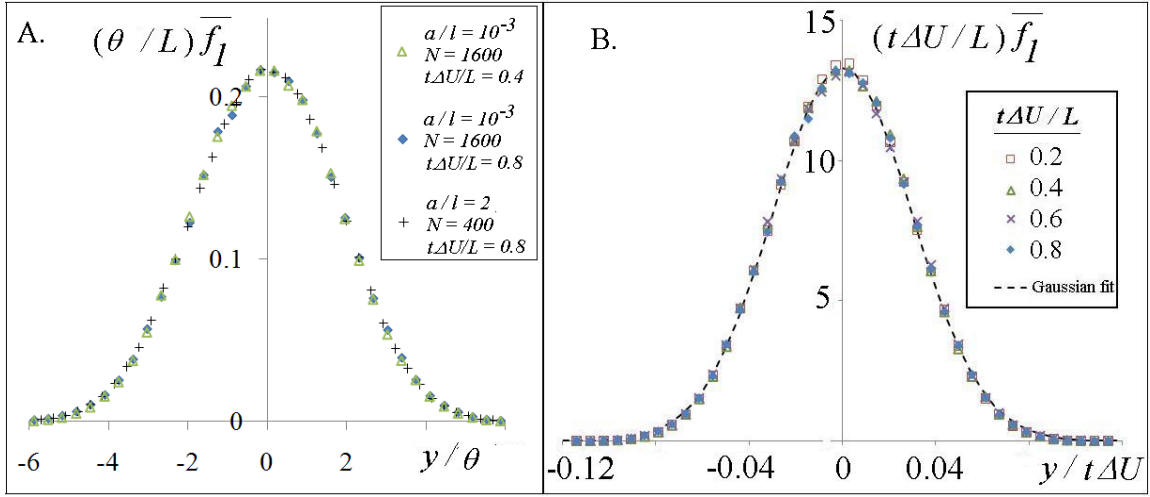

**Figure 9.** A. Single particle distribution function for different cases when scaled by momentum thickness.
B. Self similar scaling of single particle distribution function (for R4-1600)

From Figure 9A, which shows Regime II data, at different times, for two cases in which initial conditions and number of vortices are both different, $\overline{f_1}[y, t]$ takes the universal form given by

$$\frac{\theta[t]}{L}\overline{f_1}[y, t] = \Phi_1\left[\frac{y}{\theta[t]}\right] \quad (13)$$

where $\Phi_1$ is the self-similarity function; i.e. a function of two independent arguments $y$ and $t$ is reduced to a function of only a function of one argument $y/\theta[t]$.

Further, when $t\Delta U$ is such that $C_3 \ll t\Delta U \ll L$, Eq.12 shows that $\theta$ is linear in $t\Delta U$, as demanded by (17). Therefore, in the limit of $t/t_{\text{RI}} \to \infty$ (equivalently, $L/l \to \infty$ for a given $a/l$), $\overline{f_1}$ follows self-similar scaling in Regime II :

$$\overline{f_1} = \frac{L}{t\Delta U}\Phi_2\left[\frac{y}{t\Delta U}\right] \quad (14)$$

This limit is closely reached for the case R4-1600 beyond $t\Delta U/L = 0.2$ as seen in Figure 9B. This result is important, for the similarity form of the solution (18) is not admitted by the kinetic theory proposed by Chavanis (Eq. 129 of (Chavanis 2001)).

To explore this issue further, we compute the two-vortex distribution function $f_2[x_1,y_1;x_2,y_2]$ by enemble-averaging the product of the number of vortices in two given boxes around $(x_1,y_1)$, $(x_2,y_2)$ at a given time. We define the two-vortex correlation function $f_2'$ as

$$f_2'[x_1,y_1,x_2,y_2] = f_2[x_1,y_1,x_2,y_2] - f_1[x_1, y_1]f_1[x_2, y_2] \qquad (15)$$

If $f_1[x_1, y_1]$ is statistically indepenent of $f_1[x_2, y_2]$, i.e. if we make the analog of Boltzmann's 'molecular' chaos assumption, then the right hand side will vanish. Now due to $x$-homogeneity $f_2'$ should depend only on $y_1, y_2$ and $|x_1$-$x_2|$ for a sufficiently large ensemble. Again averaging over $x$ to improve the statistics we present $\overline{f_2'}[|x_1 - x_2|, y_1, y_2]$ versus $|x_1 - x_2|$ and fixed $y_1$ and $y_2$ (both set close to zero). It can be immediately seen from Figure 11 that $f_2'$ shows a systematic variation with $r$ and that it can take values several times that of the local $f_1 * f_1$ at small $|x_1 - x_2|$. Furthermore $\overline{f_2'}$ takes both positive and negative values, indicating the presence of strong two-vortex correlations of both signs alternating between each other.

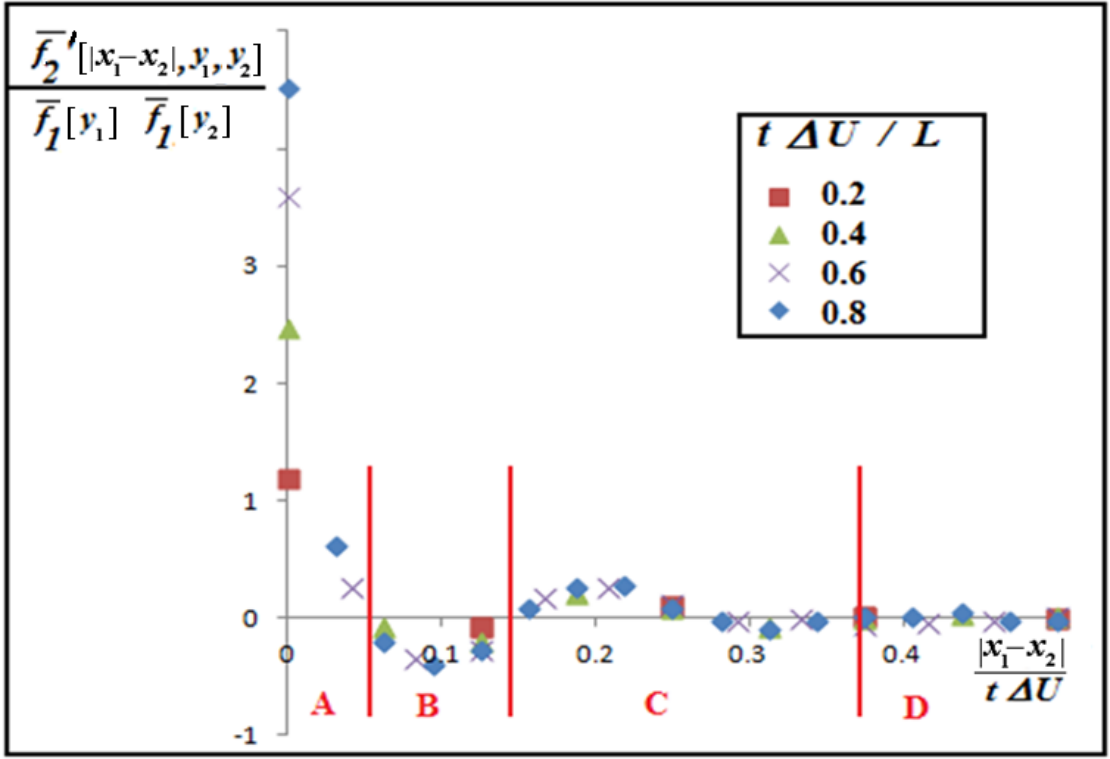

**Figure 10.** Temporal evolution of $f_2'$ as a function of $x$-separation ($r$) at $y_1 = y_2 = 0.0029\ t\Delta U$ during Regime II of case R4-1600. Note that there is self-similar scaling except at very small $r$ (region A).

To understand the $r$-dependence of $f_2'$ exhibited in Figure 10, it is instructive to relate it to the coherent structures in the flow, in particular to the length-scales associated with their size and the spacing. These are obtained as follows. From an analysis of the snapshots of the vortex configurations such as those in Figures 2 and 5, we find that the average number of coherent structures in the $L$-domain during Regime II is approximately $4L/t\Delta U$ in the limit $C_3 \ll t\Delta U \ll L$ and hence the average $x$-distance between their centers is approximately $(1/4)\ t\Delta U$, equivalently about $15\theta$ from Eq.12. Further, from Figure 2, the size of the structures is approximately half the spacing between their centers, i.e. about $(1/8)\ t\Delta U$ or $7.5\ \theta$. The nearest vortex-sparse region and therefore lies between approximately $0.06$ and $0.19\ t\Delta U$ from the center of a structure.

Returning to Figure 10, it is seen that the functional dependence of $f_2'$ on the $x$-separation exhibits four distinct regions.

A. At small separations ( $|x_1 - x_2| < 0.05\ t\Delta U$ near the x-axis), which approximately correspond to distances within the same structure (i.e. less than half the average size of the structure), $f_2'$ is several times $f_1 * f_1$ and positive.

B. At distances $0.05\ t\Delta U < |x_1 - x_2| < 0.15\ t\Delta U$, $\overline{f_2'}$ is of order $f_1^2$ and negative. This clearly characerizes the vortex-sparse region between two neighbouring structures.

C. At somewhat larger separations $f_2'$ oscillates between positive and negative values, with amplitude diminishing with distance. The first positive peak is located at approximately $0.22\ t\Delta U$, which is roughly the distance to the center of the next structure, and reflects the degree of order in the arrangement of nearby structures. The peaks progressively decay with larger separation.

D. At large distances ( $|x_1 - x_2| \geq 0.4\ t\Delta U$ ) $\overline{f_2'}$ is negligible, indicating that vortex positions are uncorrleated. It is only in this region that the analog of Boltzmann's 'molecular chaos' is valid.

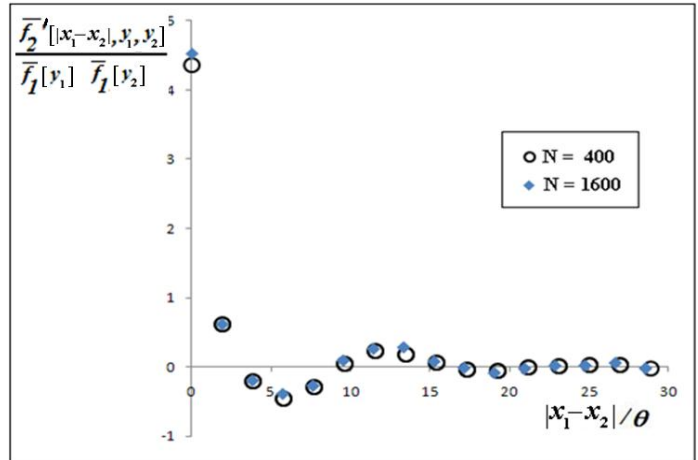

**Figure.11.** Variation of $f_2'$ with number of vortices (case R4-400 and R4-1600 at $t\Delta U/L = 0.8$ with $y_1 = y_2 = 0.18\theta$). Note that the maximum value of $f_2'$ changes by less than 5% from N = 400 to 1600 (which is within the statistical uncertainity).

We note from Figure 11, which compares $f_2'$ (at $y_1 = y_2 = 0.18\theta$, $t\Delta U/L = 0.8$) computed from simulations with N = 400 and N = 1600, that the observed values of $f_2'$ seem to be N independent. Hence in the vortex gas mixing layer, $f_2'$ can neither be neglected as done in most Boltzmann inspired 'kintetic theories' (Chavanis 2001,2005,2010, Sano 2008) nor be considered as O(1/N) as proposed in a recent work (Chavanis,2011]. This indicates that large values of $f_2'$ in the present system are not finite-N effects (as in the analysis of Chavanis), but rather indications of presence of strong correlations. This is possibly due to the nature of vortices to cluster and form coherent structures, since the nature of variation of $f_2'$ are observed to be related to the observed size and spacing of the coherent structures.

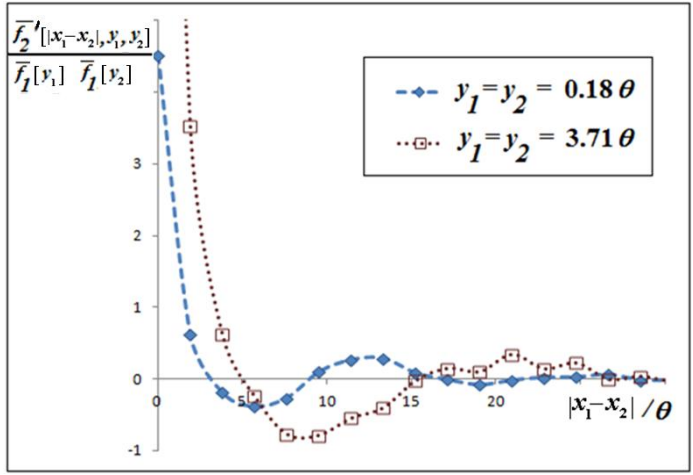

**Figure 12.** Variation of $f_2'$ with $y$ at $t\Delta U/L = 0.8$ for case R4-1600.

Figure 12 shows compares $f_2'$ at different values of $y$ co-ordinates. It can be seen that the variation of $f_2'$ is qualitatively similar by quantitatively different at different values of the $y$ co-ordinate. This may be related to the scatter of $y$ locations of the coherent strcutures, but a detailed analysis of complete structure of $f_2'$ is to appear as a separate publication.

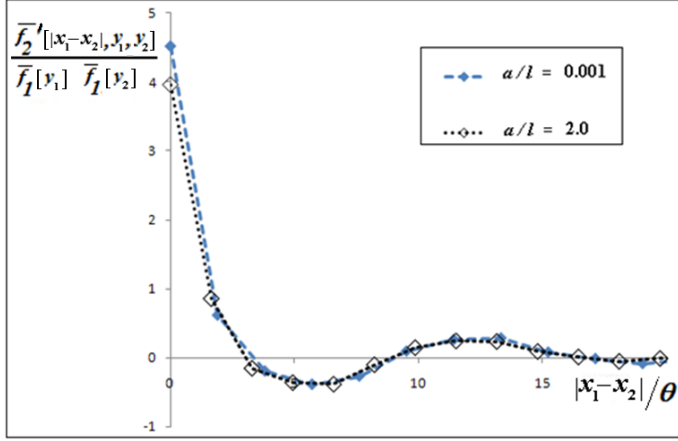

**Figure.13.** Dependence of $f_2'$ on initial conditions ($t\Delta U/L = 0.8$)

Figure 13 shows the variation of $f_2'$ for two initial conditions whose amplitudes differ by three orders of magnitude. The differences are negligible in general, but become barely noticeable at $|x_1 - x_2| \to 0$. On returning to Figure 11, we can also observe the lack of self similarity $f_2'$ of as $|x_1 - x_2| \to 0$. However, there does not seem to be any evidence against self-similarity and universality of $f_1$ (Figure 9).

A possible heuristic explanation for this apparent inconsistency is as follows. When the structures grow in size with time, the average inter-vortex spacing is expected to increase in most parts of the system. But this has to be balanced by the reduction of inter-vortex spacing somewhere in the system, possibly near the center of the structures, as demanded by the conservation of the Hamiltonian. This explanation would not be inconsistent with the observation self-similarity of $f_2'$ except at the center of the structures, i.e. when $|x_1 - x_2| \to 0$, where $f_2'$ increases with time (i.e. the vortex density and hence the correlations increase at the center of the structures as they grow in size with time). Similarly, different class of initial conditions would have a different Hamiltonian that will be conserved throughout the evolution, and this dependence is likely to manifest at the center of the coherent structures. This dependence possibly appears as the slight difference in $f_2'$ at $|x_1 - x_2| \to 0$, for the two different initial conditions shown in Figure 14. However, it has to be noted that the coherent structures occur at different *y*-locations, as observed in Figure 2. As a consequence, on averaging over different realizations, the effect of the vortex distribution within the clusters plays an insignificant role in determining the single particle distribution function $f_1$. This will be explained in detail during the analysis of Regime III in the following section. We shall show that as long as more than one structure is present (which is the case in Regime II), the vortex distribution within the structure does not alter the single particle distribution function and hence the mean vorticity and velocity profiles, and thickness, but becomes important when only one structure is left in the domain. )

The central message these analyses reveal is that the present system of the vortex gas mixing layer is strongly correlated and existing 'kinetic theory' approaches based on the Boltzmann equation that neglect correlations or consider them as $O(1/N)$ are inapplicable in Regime II. (However, several of the features of the two particle correlation function can be qualitatively explained by relating them to the coherent structures.)

## 6. REGIME III AND CONNECTIONS TO EQUILIBRIUM

Most statistical-mechanical analyses of a system involve questions regarding its final or asymptotic state. For an isolated vortex gas system, it was proposed by Onsager that for energies greater than a critical value, the formally defined temperature becomes negative and leads to emergence of large scale order by clumping of like-signed vortices. These ideas were further developed by Lundgren & Pointin (1977), who derived a closed form expression for the equilibrium single-particle distribution of point vortices of same sign and identical strength in an infinite plane. Miller (1990) and Robert & Sommeria (1991) developed statistical theories for Euler equations. The latter theory was applied to determine the final state of a temporal mixing layer in an Euler fluid (Sommeria et al, 1991) and compared with 2D Navier-Stokes simulations of the same flow.

The kinetic theory of vortices formulated by Chavanis (2001, 2010), in a spirit similar to that of the Boltzmann equation for gases, considers the evolution of the system as a relaxation to a 'Boltzmann distribution' defined for the vortex gas. As mentioned in section 5, Chavanis considers correlations to be $O(1/N)$ and hence negligible as $N \to \infty$. In more recent work (Chavanis 2011) the relaxation to equilibrium is proposed to have two stages – a 'violent' relaxation driven by Euler dynamics to a non-Boltzmann quasi-stationary state described by the Miller-Robert-Sommeria theory (which is a fluid dynamic analog of the Lynden-Bell theory (1967) of stellar systems), followed thereafter by a 'slow' relaxation driven by finite-*N* effects to the Boltzmann distribution.

The present simulations provide a basis for an assessment of the various theoretical ideas concerning relaxation an asymptotic state. Figure 14 shows the evolution of momentum thickness during long-time vortex-gas simulations in six cases. We recall from section 3 that Regime III is that part of the solution where the statistics depart from the universal linear growth of Regime II and exhibit a dependence on (and scale with) domain size. Figure 15 reveals that this Regime has three distinct sub-regimes. It can be observed in all the cases simulated here that, immediately following Regime II, there is a rapid but non-linear and domain-dependent increase in thickness that scales with *L* and extends to about $t\Delta U/L \sim 4$. We shall call this regime III(a). Beyond this, the thickness evolves very slowly over very long timescales changing over $10^4 L/\Delta U$ by less than 20% of its value at 4 $L/\Delta U$. This sub-regime is labeled III(b), and appears to reach asymptotically a final state of constant thickness, Regime III(c). The endpoint of Regime II is at $(t-t_o)\Delta U/L \approx 2$, and coincides with the beginning of III(a). The boundary between Regime I and II is highly variable, and depends strongly on initial conditions.

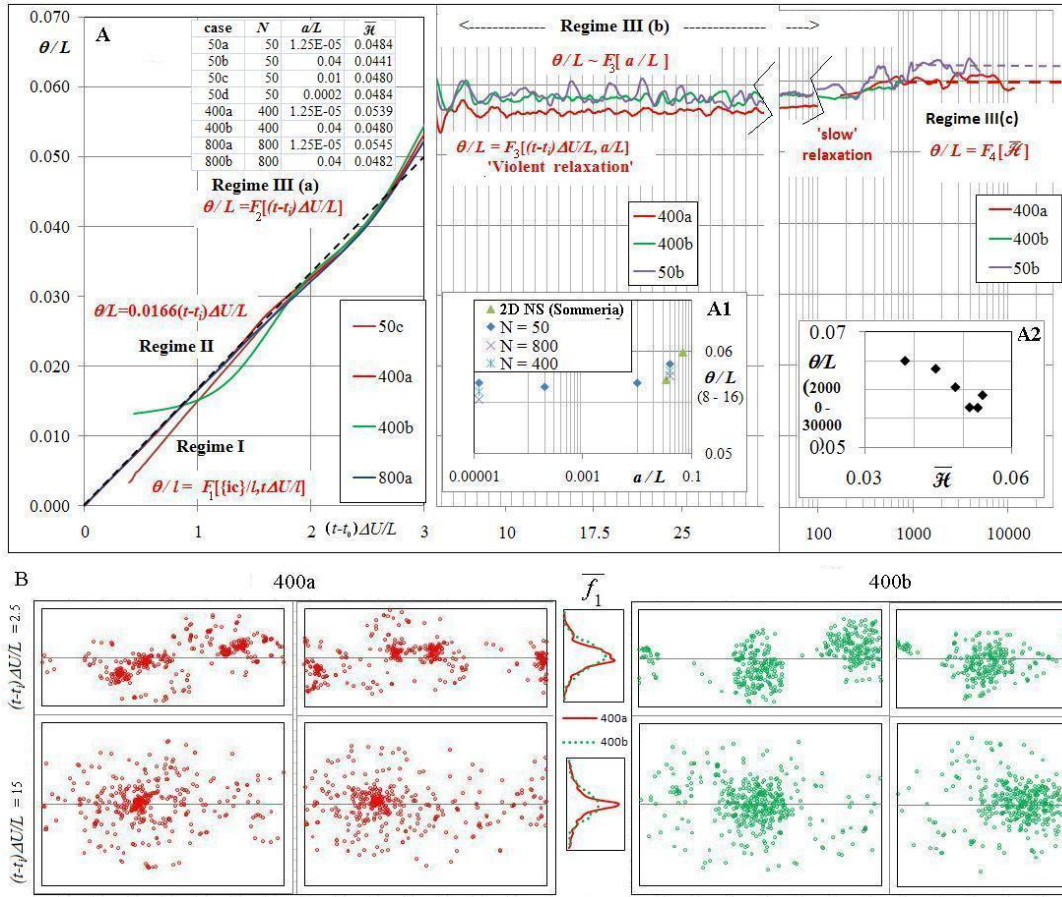

**Figure.14A**. Complete evolution of mixing layer thickness, showing sub-regimes of Regime III and their relation to Regimes I and II. **B**. Distribution of vortices within 'structures' in Regimes III(a) (top row) and III(b) (bottom row), with 'quiet' (case 400a) and 'highly disturbed' (case 400b) initial conditions.

Each of these regimes is now considered in turn.

Regime III (a)

From Figure 14A it is seen that the evolution of momentum thickness begins to depart from the linear growth of Regime II at around $(t-t_o)\Delta U/L \sim 1$. However the variation of $\theta$ continues to remain universal and independent of initial conditions or $N$ with $L$ as the length scale, as illustrated in Figure 15A for four widely different cases, till about $t\Delta U/L \sim 3$. We find that in this regime (III(a)) the number of structures can vary from 4 to 1.

Figure 14B shows a snapshot of vortex locations for two cases 400a, 400b (with two different realizations in each case), both having the same number (400) of vortices but initial $y$-displacements drawn from uniform random distributions of widely different amplitudes, $1.25 \times 10^{-5} L$ (to be called the 'gentle' case) and $0.04\ L$ ('highly disturbed' stronger case) respectively. It can be observed from Figure 15B that in the left panel, simulations in the quiet case show structures with dense cores, characterized by high concentration of vortices. With highly initial disturbances (right panel, 400b) the cores are not so dense.

From the snapshots in the top panel in Figure 15B, at $t\Delta U/L = 2.5$ corresponding to an early phase in Regime III(a), it can be seen that the size and relative locations of the two coherent structures present in the domain are similar. Also, there is almost no difference in the $x$- (and ensemble) averaged single-vortex distribution function $\overline{f}_1$. This suggests that the averaged vorticity and velocity profiles have a much stronger dependence on the distribution of the coherent structures in the domain than on the distribution of vortices within each coherent structure. In contrast the bottom panel, at $t\Delta U/L = 15$ in Regime III(b), shows significant differences in $\overline{f}_1$, which has a tall narrow peak at the centre, showing the small dense cores in the gentle i.c. case. These observations indicate why the evolution of $\theta$ obtained from the $x$- and ensemble averaged velocity distribution is universal for different initial condition classes, whereas the vorticity distribution within a single coherent structure is not. Variations in the $y$-locations of the structures averages out the effect of vorticity distribution with each structure. This also explains the universality of $\overline{f}_1$ in spite of the non-universality of $f_2$' at small $r$ in Regime II shown in Figure 13.

Regime III(b)

Figure 15 shows that, following III(a), the momentum thickness varies very slowly: thus $(\Delta(\theta/L) < 0.01$ during $4 < t\Delta U/L < 10^4$, a change less than 20% of that seen during $0 < t\Delta U/L < 4$); indeed, $\theta$ seems to asymptotically approach a constant value. Further, beyond $t\Delta U/L \sim 4$, there is only one structure left in the domain (for evidence on $N$ see inset in Figure 5, also Figure 15B); and the evolution of momentum thickness is no longer universal. We label this sub-regime III(b).

The lack of universality in Regime III(b) (and subsequently also in III(c)) is consistent with the argument presented above for Regime III(a), namely, the gentler initial conditions lead to higher vortex density in the core of the structure. Since III(b) involves a single structure, and since the $y$-centroid is invariant in time, the ordinate of the core of the structure would be similar in different realizations. Hence, unlike in Regimes II and III(a), an altered distribution of vorticity within the structure does affect the ensemble averaged statistics (Figure 15B). This explanation is consistent with the observation of lower thickness for the gentler initial conditions in Regime III(b), shown in Figure 15A. Further, the early part of III(b) follows Euler dynamics under certain limits and the thickness is a function of only $a/L$ as shown in the inset. This will be discussed in detail when we compare the present results to Navier-Stokes solutions in Section 7.

Regime III(c)

We label the statistically stationary asymptotic state ($t\Delta U/L \to \infty$) as III(c). In the present simulations with $N = 400$, this is practically reached at $t\Delta U/L > 10^4$.

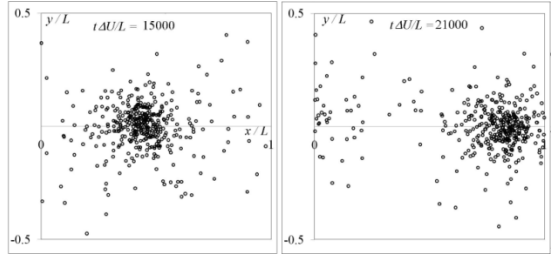

**Figure.15.** Snapshots of the lone vortex structure at two different times in III(c).

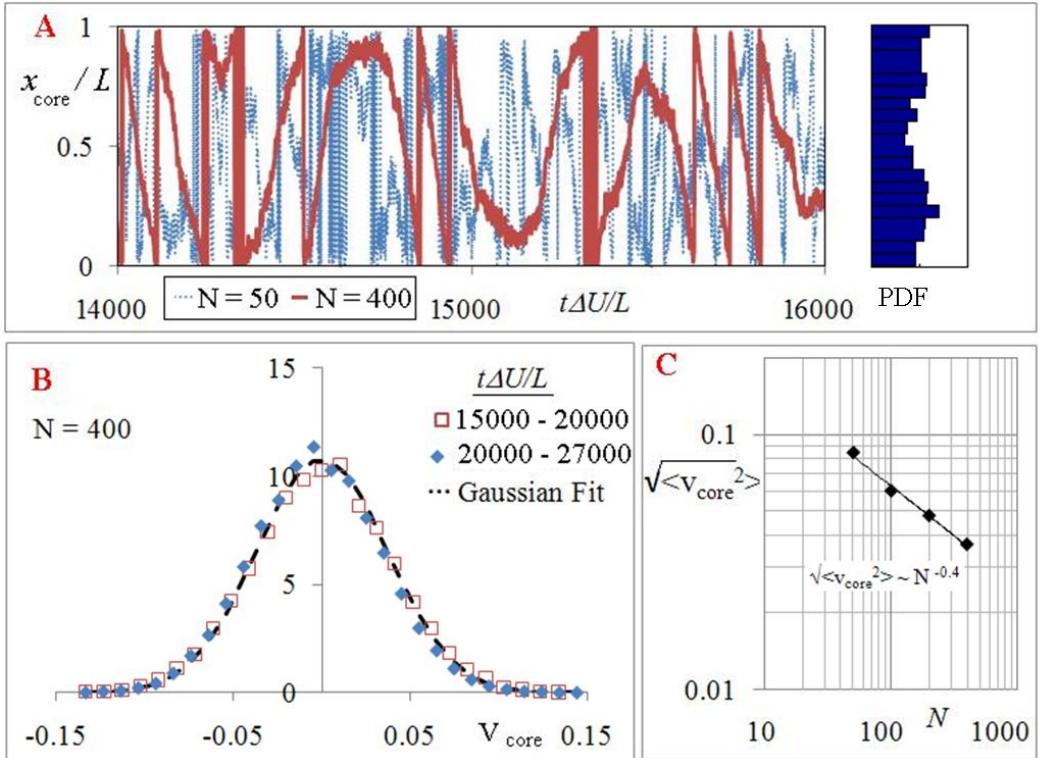

**Figure.16.A**. Motion of the core of the structure in Regime III(b). The solution samples all $x$-translated states but with very long timescales that increase with increase in number of vortices. Also shown is a histogram of the locations sampled by the core over $t = 15000 – 27000\ L/\Delta U$ for $N = 400$. **B**. The motion of the core relaxes to a stationary stochastic process as can been seen from the PDF of a characteristic velocity (defined as the distance moved by the core as a fraction of the domain during one $L/\Delta U$). **C**. The velocity of the core decreases with increasing number of vortices as $N^{-0.4}$.

To study this regime, four very long-time simulations ($t\Delta U/L$ up to $3\times10^4$) have been performed, with $N$ ranging from 50 to 400. Figure 15 shows snapshots of the vortex locations at two different times, respectively 15000 and 21000 $L/\Delta U$. It can be seen that at both times there is a single structure with a similar configuration of vortices within the structure, but the structure itself is found at different $x$-locations. The lone structure in the domain in fact keeps moving back and forth in $x$, sampling the entire domain over timescales of $O(10^3\ L/\Delta U)$. Figure 16A shows the time series of the position of the core of the structure. (The core represents the zone of highest vortex density, and its position is taken as the $x$-location $x_{core}$ of the vertical strip with the highest number of vortices, out of 101 vertical strips of equal width over the domain). On the right in Fig 16A the PDF of $x_{core}$ shows that it samples the entire domain with roughly equal probability. This is consistent with ergodicity because, unlike in the case of the infinite plane, the $x$-centroid (in the sense it has been used in this paper) of the present $x$-periodic system is not conserved (see Appendix C). From Figure 16B it is seen that the PDF of the velocity of the core does not change with time beyond 15000 $L/\Delta U$, and appears to agree well with a Gaussian distribution with zero mean. All this evidence confirms that $x_{core}(t)$ tends to a stationary stochastic process in the limit $t \to \infty$ (at each $N$).

As the size of the structure in the final state scales with $L$, the effect of boundaries cannot be neglected in III(c), however large the domain may be. This shows that the common argument about independence from the boundary, widely used in much of statistical mechanics, is not applicable to describe the final state of the present system involving long range interactions.

However, according to Figure 16A the number of crossings of $x = 0.5 L$ with $N = 400$ is roughly half that at $N = 50$, so the time taken by the structure for crossing the $L$-domain is twice as long at the higher $N$. From Figure 16C, the standard deviation of a characteristic velocity of the structure decreases like $N^{-0.4}$. Thus, if the limit $N \to \infty$ is taken first, the possibility that the structure may be stationary as $t \to \infty$ cannot be ruled out. The final asymptotic state, and ergodic behaviour in $x$, could therefore depend on the order in which the limits $t \to \infty$ and $N \to \infty$ are taken (as pointed out by Chavanis (2011) in the more generalized context of vortex gas statistical mechanics).

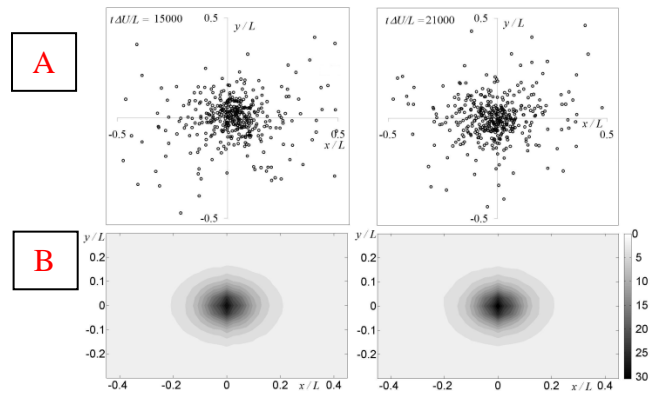

**Figure 17.A**. Vortex positions relative to (moving) centre in Regime III(b), at times $t\Delta U/L = 15000$ (left) and 21000 (right) (same data as in Figure 16). **B**. The single vortex distribution of ($x$-$x_{core}$, $y$) at $t\Delta U/L = 15000$ and 21000, averaged over 250 $L/\Delta U$). Note invariance with time.

Reverting to the limit $t \to \infty$ of an $N$-vortex system, we now study the distribution of the vortices within the structure relative to the (moving) centre. The top panel in Figure 17 shows snapshots of the positions of vortices, re-centered around the core at each of two different times separated by 6000 $L/\Delta U$. We carry out a 'short-time average' (over a duration of 250 $L/\Delta U$) of the location of vortices around the core, and these are shown in the lower panel. It is seen that there is very little variation between the single-vortex distributions across the two times. Further, as the orientation of the non-circular shape does not change, the structure as a whole is not in solid body rotation, but the individual vortices are in relative motion with respect to each other, as for example in density wave motion in galaxies.

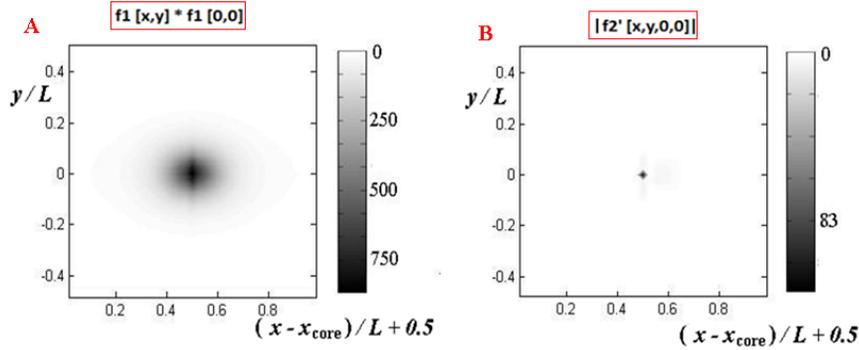

**Figure.18** Comparison of distribution functions for vortices relative to the centre in Regime III(c) (averaged over $t\Delta U/L = 27000 - 36000$). Note different scales.

This leads to perhaps the most important question in the statistical-mechanical analysis of any system, namely whether a state of equilibrium (possibly relative to a moving structure as in the present case) exists, and if so whether it has been reached or not. A necessary but not sufficient condition for the system to be in equilibrium is that *molecular chaos* should have set in; more precisely, the two-point correlation functions $f_2'$ must vanish in the thermodynamic limit $N \to \infty$. That would be consistent with all higher order correlation functions also tending to zero. This question is specifically addressed in Figure 18. It is seen that $f_2'$ computed from the time-averaged statistics in the frame of reference of the moving structure (Figure 18B) is small compared to $f_1 * f_1$ (Figure 18A). Thus molecular chaos ($f_2 = f_1 * f_1$) might be a reasonable assumption in attempts to analyze the statistics of the distribution of vortices within the moving structure. A comparison may be made with the same distribution function in Regime II, where the difference $f_2 - f_1 * f_1$ was found to be significant (Figure 10).

Now we turn to the necessary and sufficient conditions for equilibrium: the single-particle distribution must be independent of time and be governed by a single parameter, namely the temperature or its equivalent. The time-independence has already been satisfactorily demonstrated in Figure 17A. The temperature may be estimated based on the results of Joyce and Montgomery (1973) and Chavanis (2001). According to them equilibrium is described by single-particle 'Boltzmann distribution' $f \sim \exp[-B\psi]$ where $B$ is proportional to inverse temperature.

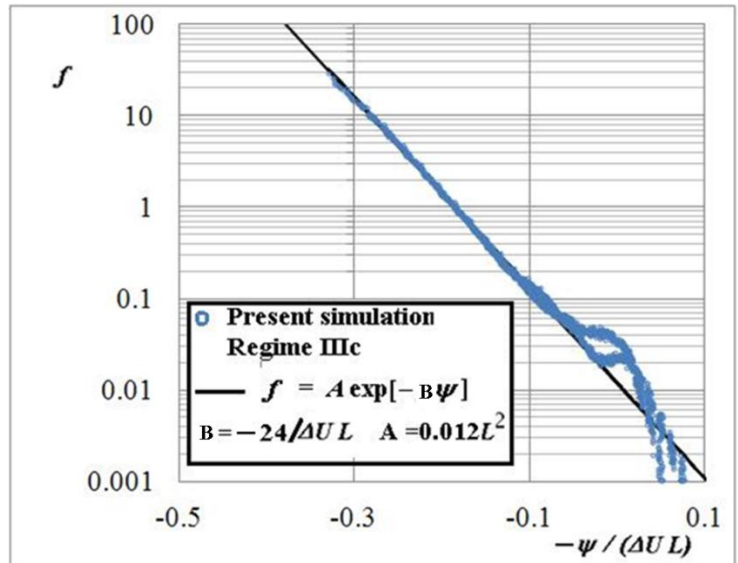

**Figure 19**. The stream function-vorticity relation in the frame of the moving structure in Regime III(c) (averaged over $t\Delta U/L = 27000 - 36000$) for case 400a.

The averaged stream function at the $i^{th}$ box can be computed numerically from the computed discrete values of $f_1$ using the expression

$$\psi(i) = L\Delta U \left(\frac{1}{4\pi}\right) \sum_{j\neq i} f_1(j) \ln\left(\frac{1}{2}\left[\cosh\left(\frac{2\pi(y_i-y_j)}{L}\right) - \cos\left(\frac{2\pi(x_i-x_j)}{L}\right)\right]\right) \Delta x\, \Delta y \quad (15b)$$

For the present simulation of case 400a, it is seen from Figure 19 that the $\psi - \omega$ relationship in Regime III(c) follows the Boltzmann distribution for $f_1 > 0.03$. The value of $B$, estimated by best fit to the time averaged data, is $-24/(\Delta U\, L)$. The temperature is clearly negative, corroborating the seminal ideas of Onsager. The two seemingly distinct branches for $f_1 < 0.03$ correspond to data from the top and bottom of the layer, implying some up-down asymmetry, but this is expected to disappear with sufficiently long averaging.

Since $\langle\omega\rangle = -\nabla^2 \psi$, a PDE can be written for the stream function (Joyce & Montgomery 1973) using as source density the Boltzmann distribution for $f_1$,

$$-\nabla^2 \psi = A \exp[-B\psi]. \quad (16)$$

A closed-form solution of the above equation was derived by Lundgren & Pointin (1977) for the infinite plane, but none has been reported so far under the present boundary conditions.

One interpretation of the present results emerges from a comparison of the distribution function obtained numerically here for the shear layer with the Lundgren-Pointin solution for the infinite plane,

$$P[r/Ro] = \frac{\tilde{A}\exp[-(1+\lambda)(r/R_0)^2]}{\left(1 - \pi\lambda\tilde{A}(r/R_0)^2\right)^2} \quad (17)$$

Here $r$ is the radial distance from the centroid, $R_0$ is the radius of gyration, $\lambda$ is proportional to the inverse temperature and $\tilde{A}$ is a normalization constant that ensures $\int_0^\infty 2\pi r P[r]dr = 1$. Some words of caution are however necessary here as the infinite plane problem studied by L-P has the Hamiltonian given by Eq.3, which is clearly different from the present Hamiltonian given by Eq.5. The conserved quantities are also different in the two cases.

A value for $\lambda$ can also be determined by making a best fit to the L-P distribution function (17). As seen in Figure 18B, the vortex distribution in the present problem is not isotropic, because of periodicity only along the $x$-direction and a domain that extends to $\pm\infty$ in $y$. To analyze the distribution we perform sector-wise averaging in the $xy$ plane, and renormalize with the number of vortices in the respective sector. We then find (Figure 20) that the radial distribution of vortices in each of the three distinct sectors shown in Figure 20 approximately follows a truncated Lundgren-Pointin distribution, with $\lambda$ (determined by best fit) taking the values $-0.972$, $-0.985$ and $-0.989$ as we move from the sector covering the $x$-axis to that covering the $y$-axis. With the use of the present non-angular description for $x$, the distribution function also gets truncated. Once again, these results are consistent with Onsager's ideas on negative temperature associated with the presence of ordered structures in the flow.

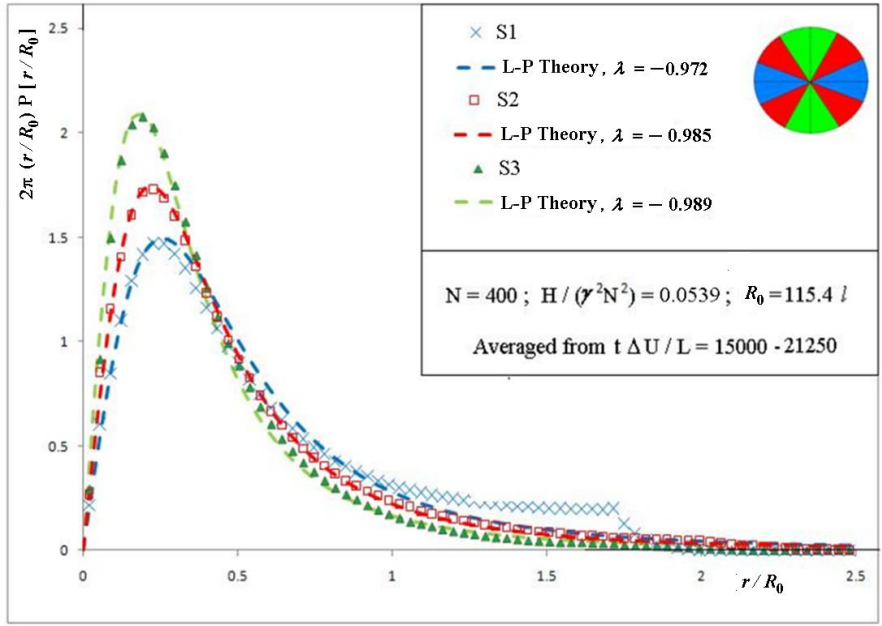

**Figure 20.** Sector averaged radial (core-centered) distribution function in Regime III. The each sectorwise distribution functions is similar to a Lundgren-Pointin equilibrium type but of different 'temperatures'. $R_0$ is the initial value of the second moment. Note that the sectors are chosen only for illustration and numerical convenience and that the 'temperature' is expected to continuously vary with $\tan^{-1}[x/y]$.

## 7. RELEVANCE TO NAVIER-STOKES MIXING LAYERS

So far we have considered the vortex gas mixing layer as a prototypical problem in non-equilibrium statistical mechanics in its own right. It is however known that under certain conditions discrete point-vortex simulations can tend to smooth solutions of the Euler equations (Beale & Majda 1983, Marchioro & Pulvirenti 1994). Also as discussed in the introduction, the effect of viscosity becomes vanishingly small at high Reynolds numbers in turbulent free shear flows. Therefore we now seek to analyze connections with 2D and 3D Navier-Stokes mixing layers.

### 7.1. Regimes I and II: Comparison with laboratory experiments

It is now well known from experiments that plane mixing layers are dominated by the largely inviscid interaction of quasi-2D coherent structures and that growth occurs through amalgamation of such coherent structures (Brown & Roshko 1974); (see in the Introduction). Vortex gas simulations show the same mechanisms in operation. It is therefore worthwhile to make some brief comparisons with 'real' (i.e. classical 3D Navier-Stokes) mixing layers. Regimes I and II are most relevant to what is observed in experimental studies of (spatial) mixing layers, and as an understanding of the two regimes is central to turbulent shear flows in general, it is particularly worthwhile to explore this issue in detail.

The explosively relaxing Regime II is analogous to the 'self-preserving' state in a turbulent shear flow, where the statistics are specified to be self-similar (Townsend, 1956). (It is paradoxical that the evolution or flow development that is labeled as 'equilibrium' flow in fluid dynamics is a highly non-equilibrium phenomenon in statistical mechanics.) Briefly, the first step in a fluid dynamical analysis of a canonical Navier-Stokes temporal mixing layer (with $L \to \infty$) would proceed as follows. From dimensional analysis,

$$\frac{d\hat{\delta}}{d(t\Delta U)} = F\left[Re, \{i.c.\}\right] \tag{18}$$

The two major assumptions hypotheses introduced at this stage: (i) any turbulent flow (subjected to constant boundary conditions) evolves asymptotically to a state independent of the detailed initial conditions excepting for any integral invariants demanded by mass, momentum and/or energy conservation, and (ii), 'if the equations and boundary conditions admit a self-preserving solution the flow asymptotically tends to that solution'. Both hypotheses, while being controversial, are extensively used in turbulent shear flow analyses (see 'working rules' (2) and (3) in Narasimha (1990)).

As $t \to \infty$, the initial conditions are assumed to be 'forgotten' andthe effect of Reynolds number becomes vanishingly small at sufficiently large Reynolds numbers. Equation (14) then reduces to

$$\frac{d\hat{\delta}}{d(t\Delta U)} = C_2 \tag{19}$$

The question of whether the constant $C_1$ is universal in the vortex-gas mixing layer is analogous to the controversy on the possible dependence of the self-preservation state of turbulent shear flows on initial conditions (George 2004, Oster & Wygnanski 1982, Balaras 2001) i.e. whether $C_2$ is a universal constant. The relation between $C_1$ and $C_2$ will be further discussed below.

The question of universal self-preservation in turbulent shear flows has been debated extensively elsewhere (Narasimha 1990, Narasimha & Prabhu 1972, Oster & Wygnanski 1982, George 2004 ..), but we shall confine ourselves here to periodically forced mixing layers, which provide excellent test cases for this purpose. One reason is that the most dominant initial perturbation is accurately known experimentally in these flows. A second is that experiments show that periodic forcing greatly alters the development of the mixing layer and this has led to strong doubts about universality (e.g. Oster & Wygnanski 1982, Ho & Huerre 1984). Periodic forcing can be imposed in many ways: oscillating the free streams (e.g. Ho & Huang, 1982), acoustic excitation by loud speakers (Husain & Hussain, 1995) or periodic deflection of a flapper at the end of the splitter plate (e.g. Oster & Wygnanski 1982, Gaster et al 1985, Naka et al 2011). The last method basically imposes a periodic deflection on a vorticity layer at its origin $x = 0$ (see Figure 1c). The analogue for the temporal vortex-gas mixing layer is to have an initial $y$-displacement of vortices that varies sinusoidally with $x$, as with cases P1 and P2 discussed in section 4.

On this basis, we compare experimental results reported by Oster & Wygnanski with the temporal vortex gas simulations which have approximately the same amplitude to wavelength ratio. Two such cases are shown in Figure 24. The comparison first involves scaling of the initial condition with wavelength, i.e. having the same value for $a/\Lambda$ in the simulation as $af/U_m$ (where, $a$ and $f$ are amplitude and frequency of flapper motion) in the experiment. Secondly it involves the Galilean transformation ($x = U_m t$). In making comparisons with wind-tunnel experiments it is necessary to allow for the presence of tunnel free-stream turbulence, which is a source of facility-specific random perturbation on the flow. We therefore add a suitable random-noise component $a_n$ to the periodic vortex deflection imposed at $t = 0$. This may also be a proxy to effects due to spatial feedback or three-dimensionality.

It is seen (Figure 21a) that periodic forcing shows results for what we shall call the short-wave case, which first enhances the growth rate, then suppresses it; finally the growth rate appears to approach the universal growth rate in Regime II. The vortex gas simulation with $a_w/\Lambda = 0.0074$ (same as in the experiments) and $a_n/a_w = 1.5$ agrees with experiments quantitatively all the way. If $a_n/a_w$ is drastically reduced to $10^{-3}$, the simulation still agrees qualitatively with the observed behaviour of the mixing layer, but has a temporal history which shows a response with a time delay of about 0.75 for $t\Delta U/\Lambda > 2$. Interestingly, addition of the noise component $a_n$ hardly affects the early evolution of the layer ($t\Delta U/\Lambda < 2$). The agreement seen in Figure 21 is therefore very encouraging for both long and short wave excitation.

These results suggest that perturbations in the initial conditions do not affect the slope $C_1$ in Eq.16 but do affect the intercept $C_3$. This is confirmed by the results shown in Figure 21b, for the long-wave case already investigated as P1 in Section 5. Here the experiments do not go beyond the initial growth-enhancement phase ($t\Delta U/\Lambda < 2$). However the simulations show excellent agreement with experiment, using very low $a_n$ ($a_n/\Lambda = 10^{-3}$), but continue into the two later phases respectively of suppressed growth and recovery towards universality, shown in Figure 21.

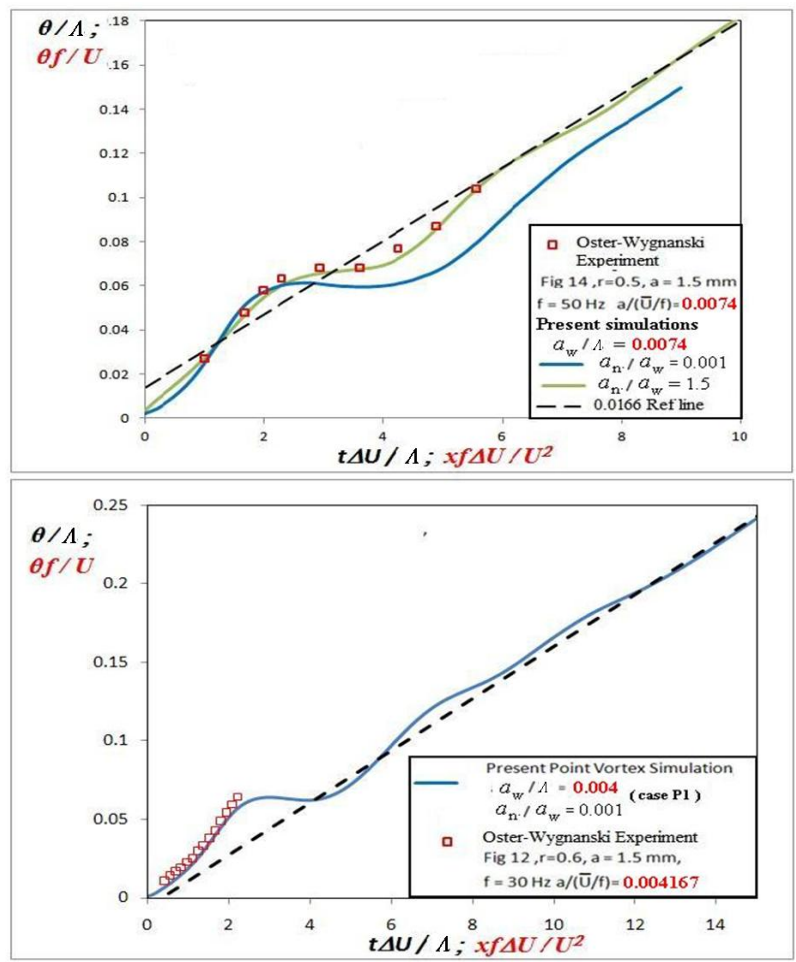

**Figure.21.** Comparison of temporal evolution of momentum thickness in the present vortex-gas computations with sinusoidal (in $x$) initial conditions ($t = 0$) with the spatial evolution of momentum thickness in experiments (Oster & Wygnanski, 1982) with sinusoidal (in $t$) forcing at $x = 0$.

The good agreement of the vortex gas simulations with the experiment over the range of data available is instructive. It suggests that the attribution of data like those in Figure 21a or b to lack of universality seems unjustified in the light of the simulations. The obvious interpretation of the experimental data is that strong periodic forcing just makes Regime I much longer. Extrapolating from the simulation results of Figure 24b on the time taken to reach Regime II, the distance necessary to reach the same state would be six times as long as the spatial range available in the experimental facility used: the wind-tunnel test section length would have to be increased from 1.5m to about 9.0m.

Further comparisons made with similar experiments reported recently by Naka et al (2010) show agreement almost as good (not shown here). In these studies also, as in Figure 21b, the entire experimental region lies in Regime I.

The Regime II universal growth rate found here is within the range quoted across different experiments (0.014 to 0.022) (see Rogers & Moser, 1994) and DNS/LES studies of 3D Navier-Stokes temporal mixing layers (0.012 to 0.018). These observations indicate that the conclusions on Regime II drawn from the present vortex gas study are relevant to real mixing layers. A more detailed analysis of the scatter in the quoted spreading rates is a more specialized fluid dynamic discussion and hence will be presented elsewhere.

**7.2. Regime III: Comparison with 2D Navier-Stokes simulations**

Regime III has, for obvious practical reasons, not been a subject of any experimental studies. However the long time 2D Navier-Stokes simulations due to Sommeria et al. (1991) are illuminating in this context and we attempt a comparison of their results with the present work. The continuum constant-vorticity layer of finite thickness (with a piecewise linear velocity profile) solved by Sommeria et al (1991) for the Navier-Stokes equations, can be accurately represented by a suitable array of point vortices in the Euler limit. One way of defining the relevant vortex gas formulation is a uniform random initial distribution with the inter-vortex spacing that is small compared to the thickness of such a layer, i.e. that $a/l \gg 1$.

As discussed in the Introduction, a 2D Navier-Stokes solution may also be expected to approach the 2D Euler solution at any finite time in the limit $Re \to \infty$. It therefore appears that both the vortex gas and the 2D NS solutions approach the 2D Euler mixing layer from different directions. It is therefore interesting to compare the solutions obtained in the present simulations with the 2D Navier-Stokes simulations of Sommeria et al (1991).

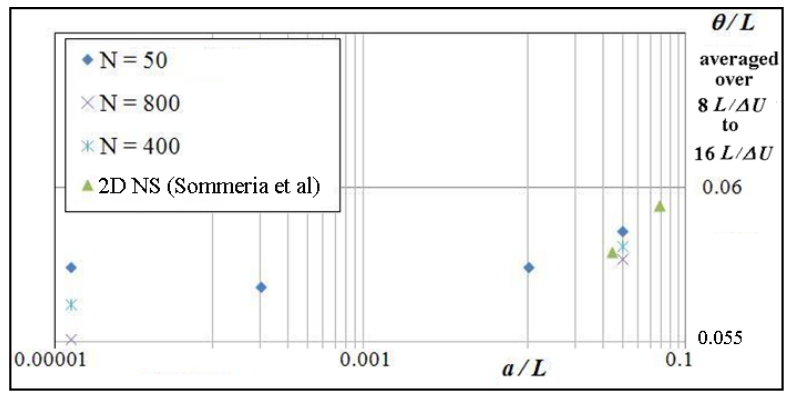

**Figure 22.** Comparison of momentum thickness in Regime III(b1) of the present vortex gas simulations with 2D Navier-Stokes (Sommeria et al, 1991)

We first return to the evolution of the momentum thickness for $t\Delta U/L < 30$ in the present vortex gas simulations, as shown in Figure 14. Sommeria et al. (1991) have studied 2D temporal Navier-Stokes mixing layers evolving from a constant vorticity band with two different thicknesses – 2 x 0.017 $L$ and 2 x 0.034 $L$ – at Reynolds numbers $L\Delta U/\nu$ ranging from 9425 to 18850 (750 – 1500 in their units). Their results show an initially rapid growth of momentum thickness, with slow changes beyond $t\Delta U/L \sim 4$. Since they do not perform ensemble averaging, rigorous estimation of spreading rates in Regime II is not possible. However, we can still compare the evolution in slowly varying Regime III.

Figure 22 compares their results with the present simulations in early Regime III(b) for the momentum thickness averaged over 8 to 16 $L/\Delta U$. At $a/L = 0.04$, the difference in averaged momentum thickness between $N = 400$ and $N = 800$ is only 1%. This suggests that for sufficiently large $N$, the solution does not depend strongly on $N$ in the vortex gas simulations, and $\theta/L$ is a function only of $a/L$. From Figure 22 it is seen that the present result is within 1% of the 2D Navier-Stokes value of Sommeria et al (case with higher $Re$ of 1500, $a/L = 0.034$).

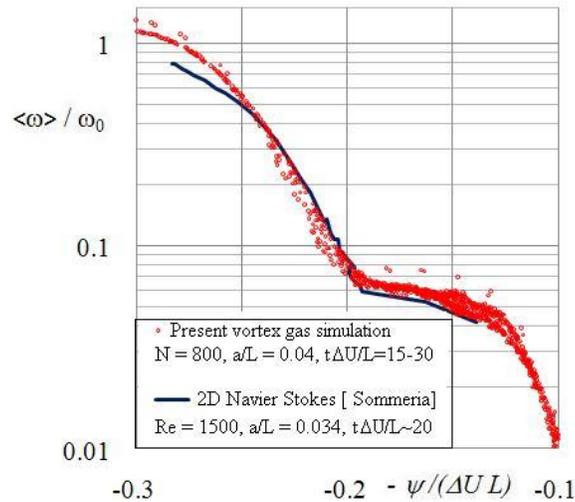

**Figure.23.** Comparison of the vorticity stream function relation between a present TMLVG computation and 2D Navier-Stokes of Sommeria et al (1991) at similar parameter values. $\langle\omega\rangle$ is the time averaged vorticity, that is proportional to $f_1$ and $\omega_0$ is the maximum vorticity in the initial condition (for the vortex gas $\omega_0$ computed from maximum value of $f_1$).

We also compare the vorticity-stream function relation in the same two simulations. As described in Section 6, the present simulations show a slowly wandering final structure in III(b). We may therefore compute ensemble and short-time averages in a frame of reference fixed with respect to the centre of the structure. This is done by dividing the domain into 101 strips of constant width in $x$ and locating the core of the structure ($x_{core}$) in the strip with the largest number of vortices. The statistics are computed using $x_i - x_{core}$ as the relevant space variable for vortex $i$ at time $t$ for each realization, and relocating vortices that have left the domain due to such an operation using the periodicity condition. Figure 23 compares that the $\psi - \omega$ relation in 2D NS calculations of Sommeria et al (1991) with the present simulations. The two agree quite closely, over the range $0.05 < \langle\omega\rangle/\omega_0 < 0.5$ at ($t\Delta U/L \sim 20$, before at times before viscous dissipation becomes dominant in the NS simulations). This agreement indicates that at such times both computations closely follow Euler dynamics.

This result must be considered of great significance, for the close agreement between two such physically and mathematically distinct approaches to the same problem establishes the quantitative relevance of the vortex gas results to high-Reynolds Navier-Stokes solutions in free turbulent shear flows.

Relation to Euler *Equilibrium*

We now turn to a discussion of the equilibrium concept. As discussed in Section 6, the statistical equilibrium for a point vortex gas (Montgomery & Joyce (1974), also see Chavanis (2001)), neglecting correlations (i.e. setting $f_2' = 0$) and taking the limit $t \to \infty$ before the limit $N \to \infty$, is shown to be characterized by the Boltzmann distribution $\exp[-B\psi]$ for $f$, where the Lagrange multiplier $B$ can be interpreted in terms of an inverse temperature. The Boltzmann distribution is obtained by maximization of $f_1 \ln[f_1]$, subject to the constraint that the Hamiltonian is conserved.

On the other hand, it has been proposed that an 'equilibrium state' for Euler flow (Robert & Sommeria, 1991) can be obtained by maximization of the Kullback entropy ($\int \{f_1 \ln[f_1] + (1 - f_1) \ln[1 - f_1]\}$ where $f_1$ is non-dimensionalized with its maximum value in the initial condition, subject to constraints of kinetic energy and linear and angular momentum of the Euler flow. This limit is expected to be reached if the limit $N \to \infty$ before the limit $t \to \infty$ (Chavanis, 2011).

The present results, shown in Figure 24, suggest that the vortex gas has a tendency to relax to the Robert-Sommeria Euler equilibrium at 'intermediate times' of $O(10\ L/\Delta U)$, and to the Boltzmann type equilibrium at much longer times of $O(10^2 - 10^4 L/\Delta U)$ for $N = 400$, $a/L = 0.04$. For the vortex gas simulations $N \ln N \sim 2400$ and $(Na/L) \ln (Na/L) \sim 45$, putting them possibly beyond the parameter range where finite $N$ effects become important. We may label these two sub-regimes of relaxation as III(b1) and III(b2).

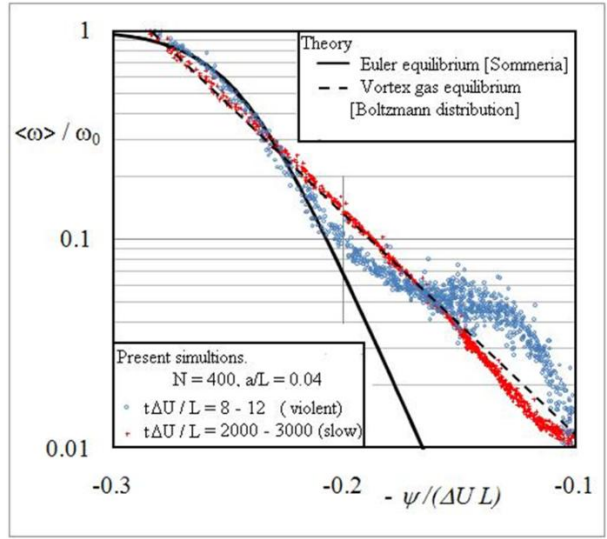

**Figure.24.** 'Comparison of present simulations at two times with Euler equilibrium of Robert & Sommeria (1991), and the 'Botlzmann' distribution of Onsager/Chavanis. Note that both the theoretical curves are two-parameter fits, and the averaging in the present simulation has been done relative to the centre of the structure.

This is not inconsistent with the recent theoretical results of Chavanis (2011), which suggest that relaxation to equilibrium in a vortex gas has two stages, namely a 'violent relaxation' that closely approximates Euler dynamics and a 'slow relaxation' driven by finite $N$ effects that appear at timescales of $O(N\log N)$, beyond which the vortex gas will relax to the 'Boltzmann' distribution. (However, the values of $N = 400$ used in these simulations are not large enough for a strict comparison with the proposed Euler equilibrium which is valid only at sufficiently large $N$, when time at which finite-N effects become important is much greater than the time required for relaxation to Euler equilibrium, or for a rigorous verification of the proposed $N\log N$ scaling.)

## 8. CONCLUDING REMARKS

In the work reported here we have uncovered, through extensive simulations, certain remarkable properties of the statistical evolution of a vortex-gas mixing layer. Among them are the existence of three distinct regimes in the evolution of the flow from a prescribed initial condition to a final asymptotic state, and their respective scaling laws. We find that evolution in Regime I depends strongly on initial conditions, Regime II exhibits a linear growth in layer thickness with time, and Regime III is a domain-dependent evolution to a final state that involves a single structure. In a simulation with 400 vortices (in the domain $0 < x < L$ with a velocity difference $\Delta U$), the asymptotic state is reached after very long times ($O(10^4 \, L/\Delta U)$) as the single structure it contains undergoes a two-stage relaxation process corresponding to what (in other contexts) have been called 'violent' and 'slow' phases. The former is driven by Euler dynamics over times of $O(10 \, L/\Delta U)$, and could be approaching the 'Euler equilibrium' described by Robert & Sommeria (1991) in their 2D Navier-Stokes solution. The slow relaxation is driven by finite-$N$ effects. These phenomenon are consistent with the recently proposed ideas of Chavanis (2011) on the approach to equilibrium in vortex gas statistical mechanics. The lone surviving structure is shown to wander randomly along the $x$-axis in the final state, uniformly sampling all $x$-translated solutions consistently with ergodic behaviour. The distribution of vortices within the moving structure can be roughly described by a truncated, anisotropic version of the Lundgren-Pointin distribution; and the vorticity-stream-function ($\psi - \omega$) relationship, approaches the Boltzmann distribution for the vortex gas, hence constituting a 'relative' equilibrium in the sense defined in Newton (2001). The $\psi - \omega$ relationship agrees with that found in 2D Navier-Stokes calculations of Sommeria et al (1991) during the early part of what we here call Regime III. It therefore appears that both the vortex gas and the 2D NS solutions approach the 2D Euler mixing layer from different directions. This suggests that the vortex gas model does provide a weak solution of the 2D Euler equations for the mixing layer.

A major finding in this work is the universality of the spreading rate ($0.0166 \pm 10^{-4}$ for the momentum thickness) of the vortex gas mixing layer in the intermediate linear growth Regime II, across a wide range of initial conditions and domain sizes. We show that the existing framework of vortex-gas kinetic theories inspired by the Boltzmann equation is not applicable in Regime II, as they do not account for the strongly correlated states created by multiple, interacting 'coherent structures' of vortices. This regime lasts for just $t\Delta U/L = O(1)$, an order of magnitude shorter and more intense than the 'violent' relaxation process that occurs in the early stages of Regime III. Regime II can therefore be justifiably called an 'explosive' phase. Regime II is the counterpart of the self-preserving flow regime observed in laboratory experiments of turbulent shear flows, in the sense that *both* mean velocity and Reynolds shear stress profiles exhibit self similarity with the *same* velocity and length scales ($\Delta U$ and $\theta$ in the present case) in the regime. Flows with such self-similarity are often said to be in a state of 'equilibrium' in the fluid-dynamical literature (following Clauser 1956). It is ironical that what represents 'equilibrium' in fluid dynamics occurs in what is an early 'explosive' phase of relaxation in non-equilibrium statistical mechanics.

From a fluid dynamics perspective, the close agreement between the evolution of momentum thickness in the present simulations for periodic initial conditions on the one hand, and the observed behavior of the mixing layer in laboratory (Navier-Stokes) experiments with equivalent periodic forcing on the other, is highly significant. The reason is that periodic forcing represents a case where the initial perturbation in the temporal simulation can be relatively precisely specified; also the vortex-gas model closely captures the observed early enhancement as well as later suppression of the growth rate relative to the unforced layer. Further, the Regime II growth rate of the vortex-gas mixing layer is within the scatter of the quoted 'self-preservation' spreading rates, across several mixing layer experiments as well as the 3D Navier-Stokes simulations. These results suggest that the momentum dispersal in the (Navier-Stokes) turbulent mixing layer is dominated by what may be called the Kelvin/Biot-Savart mechanism at high Reynolds numbers. In this approach all the dynamics is condensed into the Kelvin theorem that proves the invariance of the circulation around a contour moving with the flow of an inviscid incompressible fluid. This enables the formulation of a vortex-gas problem in which each vortex has a strength which remains invariant throughout the flow evolution. Once this is done the motion of the vortices, and consequently also the ('induced', better called 'associated') velocity field, is determined by the purely kinematic Biot-Savart relation. Importantly, the vortex gas description assumes that the flow is strictly 2D. This is not unduly restrictive for determining layer growth, as the coherent structures in plane mixing layers tend to be quasi-2D. It has been shown (Corcos &Sherman 1984, Corcocs & Lin 1984), via non-linear calculations, that three-dimensional instabilities growing side by side with 2D coherent-structure amalgamations are relatively slow-growing, as they are inhibited by the growth of the 2D instabilities. The present broad agreement is therefore consistent with the conclusion of Corcos et al. that 3D motion plays a secondary role relative to the 2D mechanisms, enabling a purely 2D model to provide a reasonable description of layer growth.

These findings suggest that the present results may be relevant to the long standing controversy of universality or otherwise of the self-preservation spreading rates of turbulent shear flows. Some experiments show that an effect of initial conditions on the spreading rate is present over the whole extent of the flow investigated (Oster & Wynanksi 1982, Oguchi & Inoue 1984, Bell & Mehta 1990, Slessor et al 1998). However one cannot, on the basis of such experiments, conclude that the effects will persist 'forever'. In any case there are also other studies which have shown that a unique self-preservation state is indeed reached after long transients. For example consider the single-stream shear layer experiments of Kleis & Hussain (1979, figure reproduced in Narasimha 1990), conducted in a 12 feet-long chamber. The authors found that the spreading rates of two mixing layers, evolving respectively from a laminar or turbulent boundary layer at the trailing edge of a splitter plate, continue to exhibit differences until approximately 5 ft., after which they both attain the same spreading rate. The present vortex gas simulations are consistent with the above example in emphasizing that the memory of initial conditions can be extraordinarily long for some initial conditions,but is finite in all cases (at least in the 2D problem). Comparison of the present simulations with the experiments of Oster & Wynanski (1982) and Naka et al (2010) suggest that, in many of their experiments, the entire flow in the apparatus may never have gone beyond what we identify here as Regime I. This result provides a part of the explanation for the scatter among the quoted 'self-preservation' spreading rates in the fluid dynamics literature. Other possible explanations of the scatter in the quoted spreading rates especially across DNS/LES studies of temporal mixing layers include insufficient averaging and estimation of (Regime II) spreading rates by fits extending to what is actually the domain affected Regime III, a detailed discussion of which will be presented elsewhere.

From a statistical mechanics point of view, many of the present results could be more generally valid for other systems with long range interactions, such as those in plasma and stellar dynamics. The most interesting statistical-mechanics finding here may be the universality of the exponent (= 1) as well as the coefficient (= 0.0166) in the growth of layer thickness $\theta = 0.0166 \, (t\Delta U)^1$ + const. with time during a highly correlated explosive relaxation that is far from statistical-mechanical equilibrium (although, ironically, a phase that is often labelled 'equilibrium' in the fluid-dynamics literature to indicate a special kind of self-similarity). While the universality of exponents is well known in the theory of critical phenomena, the universality of the multiplying coefficient found in the present system over a vast range of initial condition classes is at the least unusual. The present results may therefore have a special significance in non-equilibrium statistical physics.

## ACKNOWLEDGEMENTS

We thank Prof Garry Brown (Princeton) and Prof Anatol Roshko (Caltech) for many rewarding and enjoyable discussions and suggestions. We also thank Prof. Uriel Frisch, Prof. Joel Sommeria, Prof N. Kumar (RRI Bangalore) for their feedback on this work. We thank Dr. Ansumali (JNCASR) for suggesting bimodal initial conditions. SS wishes to thank Anubhab Roy and Dr. Saurab Diwan (JNCASR). We are grateful to Drs. S.D. Sherlekar and R.K. Lagu for providing supercomputing resources at the Tata EKA and later at Intel. We also thank Mr. Sapre (TCS) for assistance in parallelizing our code. We acknowledge support from DRDO through the project RN/DRDO/4124 and from Intel through project RN/INTEL/ 4288.

# APPENDIX A: Conserved quantities

For the present x-periodic system, the Hamiltonian is given by

$$\mathcal{H} = -\frac{\gamma^2}{8\pi} \sum_{i=1}^{N} \sum_{j=1; j \neq i}^{N} \ln\left(\frac{1}{2}\left[\cosh(2\pi(y_i - y_j)/L) - \cos(2\pi(x_i - x_j)/L)\right]\right) \quad (A1).$$

$$\frac{d\mathcal{H}}{dt} = \sum_i \left(\frac{\partial \mathcal{H}}{\partial x_i}\frac{dx_i}{dt} + \frac{\partial \mathcal{H}}{\partial y_i}\frac{dy}{dt}\right) = \Gamma \sum_i \left(-\frac{dy_i}{dt}\frac{dx_i}{dt} + \frac{dx_i}{dt}\frac{dy_i}{dt}\right) = 0 \quad (A2)$$

Consider an infinitesimal translation by $\{A_x, A_y\}$ and rotation by an infinitesimal angle $\Theta$ of the system

The $x$ and $y$ displacements of the $i^{th}$ vortex are then given by

$$\Delta x_i = A_x - \Theta y_i, \qquad \Delta y_i = A_y + \Theta x_i \quad (A3)$$

The resulting change in the Hamiltonian is given by

$$\Delta \mathcal{H} = \sum_i \left(\frac{\partial \mathcal{H}}{\partial x_i}\Delta x_i + \frac{\partial \mathcal{H}}{\partial y_i}\Delta y_i\right) \quad (A4)$$

Using Hamilton's equations (5), and substituting (A3) in (A4) we get

$$\Delta \mathcal{H} = \Gamma \left(-A_x \frac{d}{dt}\sum_i y_i + A_y \frac{d}{dt}\sum_i x_i + \Theta \frac{d}{dt}\sum_i \left(\frac{x_i^2 + y_i^2}{2}\right)\right) \quad (A5)$$

Explicit evaluation of $\Delta H$ by substitution of (A3) in (A1) gives

$$\Delta \mathcal{H} = -\frac{\Gamma^2 \Theta}{8L} \sum_{i=1}^{N} \sum_{j=1; j \neq i}^{N} \left\{\frac{(x_i - x_j)\sinh(2\pi(y_i - y_j)/L) - (y_i - y_j)\sin(2\pi(x_i - x_j)/L)}{\cosh(2\pi(y_i - y_j)/L) - \cos(2\pi(x_i - x_j)/L)}\right\} \quad (A6)$$

Comparing coefficients of $A_x$, $A_y$ and $\Theta$ in equations (A5) and (A6), we get the conservations laws for the centroids as

$$\frac{d}{dt}\Sigma_i x_i = 0 \quad (A7)$$

$$\frac{d}{dt}\Sigma_i y_i = 0, \quad (A8)$$

and the rate of change of the second moment as

$$\frac{d}{dt}\Sigma_i(x_i^2 + y_i^2) = -\frac{\Gamma}{4L}\sum_{i=1}^{N}\sum_{j=1; j \neq i}^{N}\left\{\frac{(x_i - x_j)\sinh(2\pi(y_i - y_j)/L) - (y_i - y_j)\sin(2\pi(x_i - x_j)/L)}{\cosh(2\pi(y_i - y_j)/L) - \cos(2\pi(x_i - x_j)/L)}\right\}$$

$$(A9)$$

Hence second moment is not conserved in the present model.

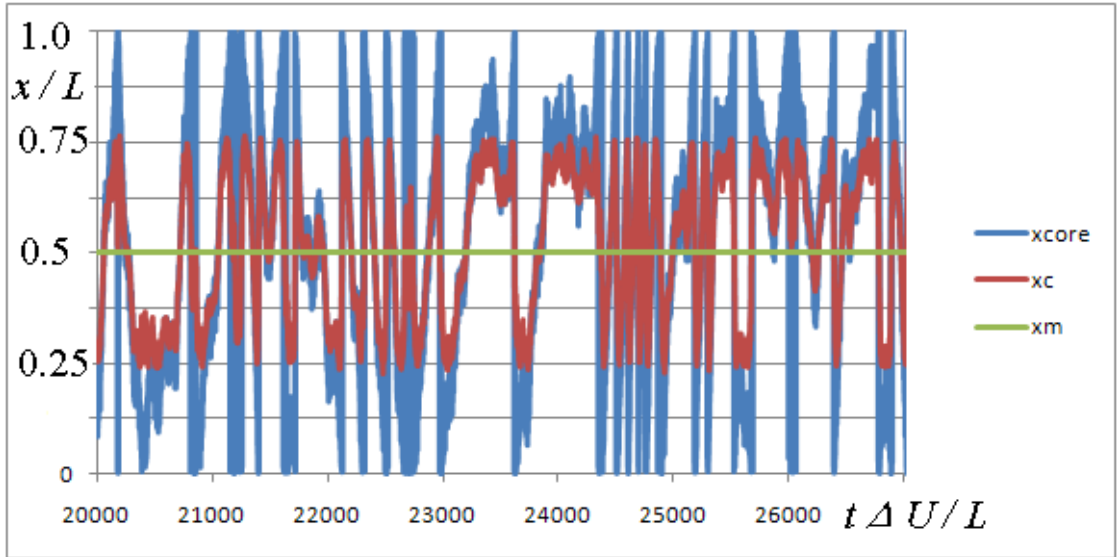

**Figure 27.** Conservation of $x_m$ and non-conservation of $x_c$, the projected centroid of the vortices in the domain.

Further, while (A8) implies conservation of the *y*-centroid, there is a subtlety in the conservation of the *x*-centroid. As the present system is periodic in $x$, $\{x_i\}$ are like angular variables, and hence $x_m = \Sigma_i x_i \bmod L$ is conserved, but the 'projected' centroid of the vortices in the domain, $x_c = \Sigma_i x_i / N$, is not conserved. This is consistent with the results of the present simulations, shown in Figure 27, where $x_c$ fluctuates between 0 and $L$ and is close to the center of the core, $x_{\text{core}}$ (except when $x_{\text{core}}$ is near $x = 0$ or $x = L$), while $x_m$ is a constant.

# APPENDIX B : Initial conditions

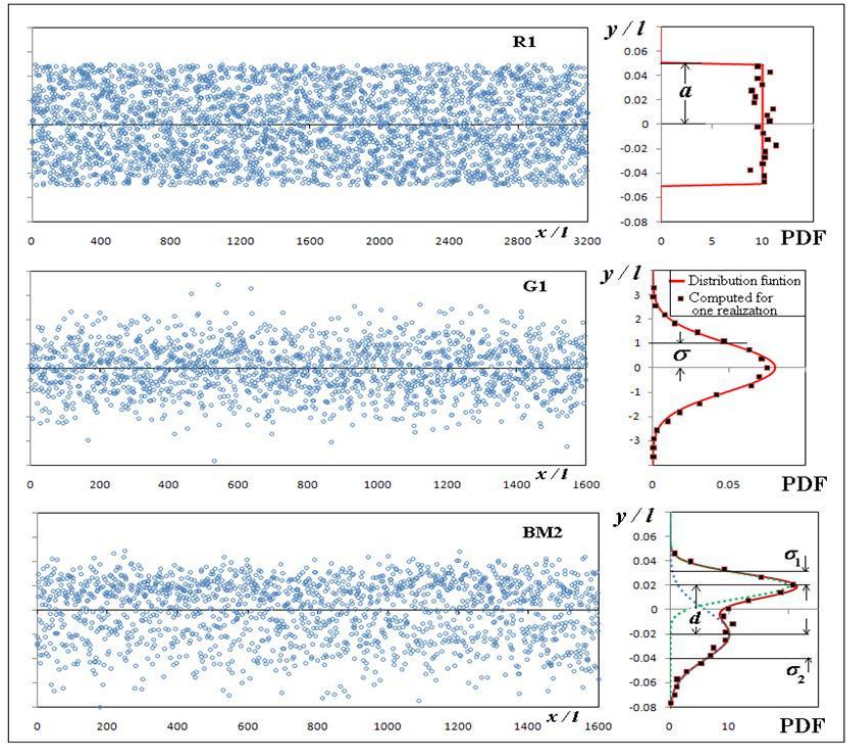

**Figure.28.** Sample Initial conditions and respective PDFs. Note that *x* and *y* are not to scale.

# APPENDIX C : Desingularization

Hama & Burke (1960) and Moore (1971) show that an increase in the number of vortices makes the system more chaotic and creates smaller scales (in the perturbation of the discretized vortex sheet). This is understandable as the Kelvin-Helmholtz instability of a vortex sheet leads to fastest growth for the smallest wavelength, which in this case is the inter-vortex spacing. Hence the system of point vortices when used as a discretization of a vortex sheet does not lead to a smooth roll-up. Acton attempts to 'cure' this difficulty by introducing a cut-off radius following Chorin and Benard (1972).

Krasny (1986) introduced a desingularized version of the governing equations, by desingularizing the kernel by adding a small positive quantity $(\epsilon/L)^2$ ($\delta^2$ in Krasny's notation) to the denominator of (1,2) ($\epsilon$ is proportional to the radius of spread of resulting vorticity field around each vortex in the units of $l$). This prevents arbitrarily large velocities close to each vortex. Hence the governing equations of such an $x$-periodic array of desingularized vortices of are given by

$$\frac{dx_i}{dt} = -\frac{\Gamma}{2L} \sum_{j=1, j \neq i}^{N} \frac{\sinh(2\pi(y_i - y_j)/L)}{\cosh(2\pi(y_i - y_j)/L) - \cos(2\pi(x_i - x_j)/L) + (\epsilon/L)^2} \quad (C1)$$

$$\frac{dy_i}{dt} = \frac{\Gamma}{2L} \sum_{j=1, j \neq i}^{N} \frac{\sin(2\pi(x_i - x_j)/L)}{\cosh(2\pi(y_i - y_j)/L) - \cos(2\pi(x_i - x_j)/L) + (\epsilon/L)^2} \quad (C2)$$

Desingularization 'filters' out the high wave number instabilities and delays the onset of chaos in the system (but as we shall show, only to a finite time depending on the value of $\epsilon$). While this method is useful to study a smooth vortex-sheet roll up, it is not useful in the present study. Further, the Hamiltonian (Eq.5) is no longer conserved, but it is possible to define an alternate Hamiltonian (C3) that will be conserved by (C1, C2).

$$\mathcal{H}_d = -\frac{\Gamma^2}{8\pi} \sum_{i=1}^{N} \sum_{j=1; j \neq i}^{N} \ln\left(\frac{1}{2}\left[\cosh\left(\frac{2\pi(y_i - y_j)}{L}\right) - \cos\left(\frac{2\pi(x_i - x_j)}{L}\right) + \left(\frac{\epsilon}{L}\right)^2\right]\right) \quad (C3)$$

The solutions of the desingularized equations have the disadvantage that they cease to be a weak solution of the Euler equations when inter-vortex distances become comparable with $\epsilon$.

Another way of understanding desingularization is that the vorticity field is no longer a set of delta functions, but is spread over a region around the center of each desingularized vortex. Hence the effective vortex sheet reprented by a row of such desingularized vortices (when the desingularization core is larger than the inter-vortex spacing) has finite thickness and hence is not unstable to small wavelength perturbations of the order of the inter-vortex spacing. But as detailed in section 2, the present system should not be considered as a representation of a vortex sheet, but rather as a vortex gas, whose chaotic evolution is not only of interest in itself but shed much light on some important aspects of the behvior of on turbulent mixing layers.

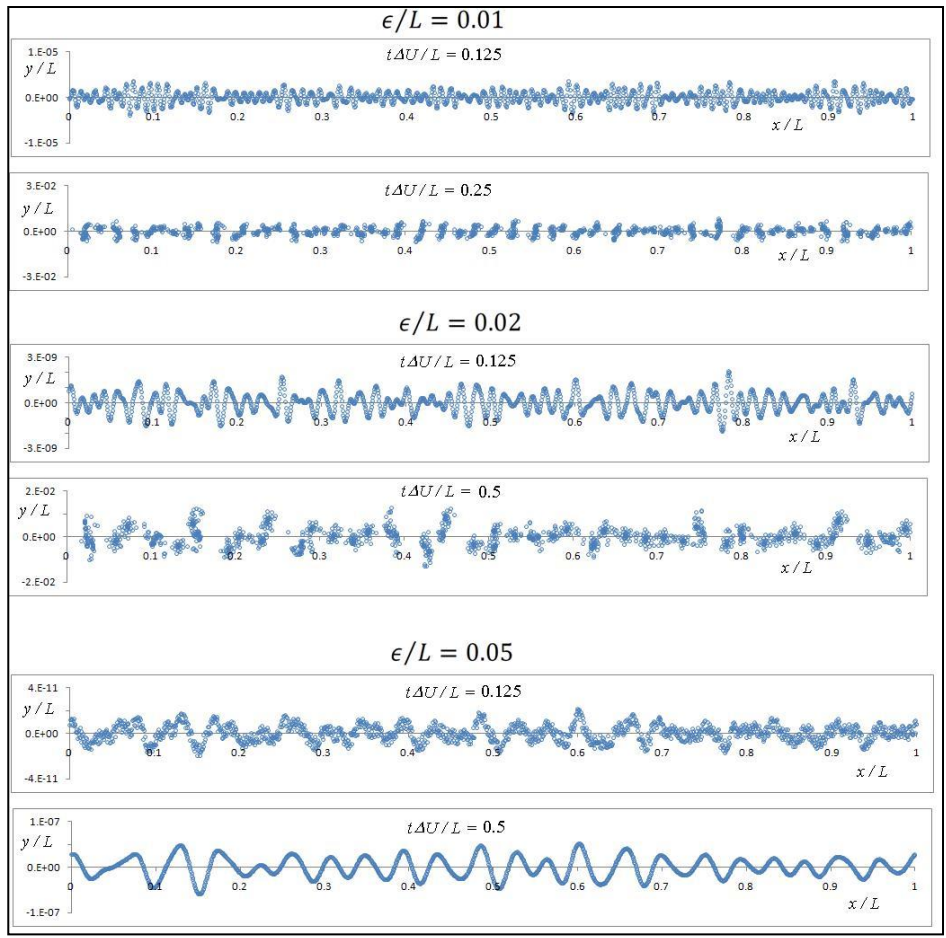

**Figure.29.** The initial development of the layer with different values on $\epsilon$. (Note that *x* and *y* are not to scale). Note that with larger values of $\epsilon$, the effective thickness (based on the velocity profile) is larger and hence the most unstable wavelength is longer and grows slower.

In order to study the effect of desingularization, simulations of case R3 are carried out for different values of the desingularization parameter $\epsilon$ and the results are shown in Figure 29A. It can be observed that the simulations with desingularization initially grow very slowly. This is because, the fast growing - short (with respect to $\epsilon$) wavelength disturbances in the random initial condition are suppressed as shown in Figure 29. However, this only increases the duration of Regime I, as once coherent structures are formed (initial size of which depends on the $\epsilon$, see Figure 29), desingularization does not prevent chaotic interaction of coherent structures.

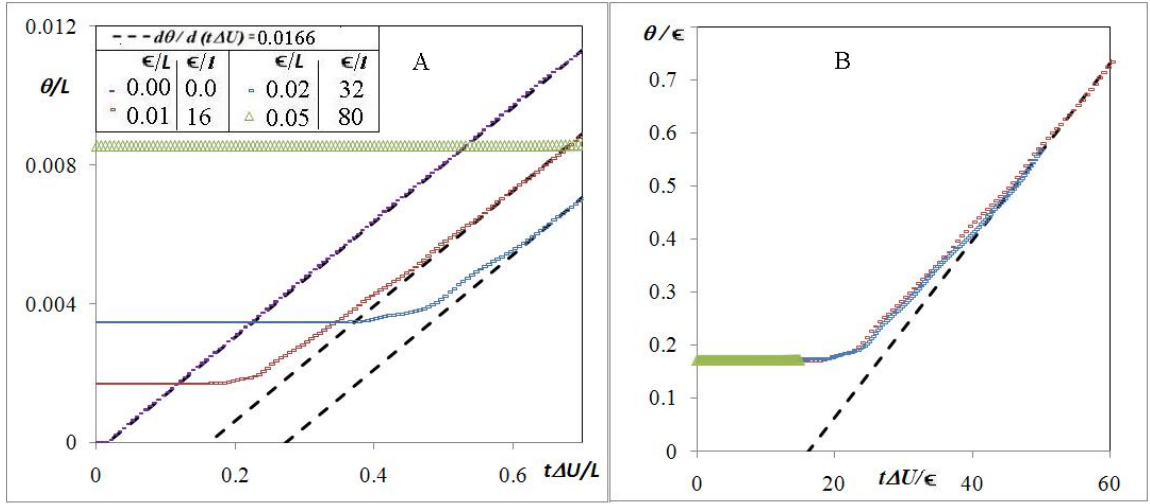

**Figure 30.** Effect of desingularization. case R3 is repeated with different values of desingularization parameter ($\epsilon$). A. Increase in $\epsilon$, while delays (in terms of $t\Delta U/L$ and $t\Delta U/l$) the onset of Regime II, has no influence on the spreading rate in Regime II, and hence the observation on universality in the non-equilibrium evolution is unaffected on desingularizing the vortices (with any given $\epsilon/l$ at sufficiently large $N$). B. Scaling based on $\epsilon$. Note that $\theta/\epsilon = F[t\Delta U/\epsilon]$ for $\epsilon \gg a$ and $t\Delta U \ll L$.

For $\epsilon = 0.01\ L$ and $\epsilon = 0.02\ L$, it is seen that the layer eventually transitions to the universal chaotic Regime II with a growth rate identical to that observed for the non-desingularized case. However, for $\epsilon = 0.05\ L$, the layer takes longer than $t\Delta U/L$ of 0.8 before onset of chaos and hence Regime II may not be very short or absent (as the onset of Regime III will begin) and is not observed within the extent of the simulation. It can be seen from Figure 30B that the evolution of momentum thickness of the different cases collapse on scaling with $\epsilon$. This is because, $\epsilon$ determines the effective thickness (based on the mean velocity profile) of the initial layer and hence the most unstable wavelength and the size and the number of the coherent structures that are initially formed at the end of the first roll up. If $\epsilon$ is of the order of $L$, only very few structures are will be formed in the periodic box and hence the layer will enter Regime III right away. But in the limit we are interested in, namely $L/l \to \infty$ for a given $\epsilon/l$, there will always be sufficient number of coherent structures left in the domain after the first roll-up and subsequent development will be dominated by chaos at the level of the coherent structures which will not be suppressed by desingularization. Hence the conclusions on universality of Regime II are unchanged by desingularization. However, the long time (Regime III(b) and beyond), where the momentum thickness was shown (in Section 6) to depend on the Hamiltonian, may be significantly different with desingularization as the Hamiltonian (5) is no longer conserved.